# The two-dimensional infinite Heisenberg classical square lattice: zero-field partition function and correlation length

Jacques Curély

Laboratoire Ondes et Matière d'Aquitaine (LOMA), UMR 5798, Université de Bordeaux
351 Cours de la Libération, 33405 Talence Cedex, France

E-mail: jacques.curely@u-bordeaux.fr

**Abstract**

We rigorously examine $2d$-square lattices composed of classical spins isotropically coupled between first-nearest neighbours. A general expression of the characteristic polynomial associated with the zero-field partition function $Z_N(0)$ is established for any lattice size. In the infinite-lattice limit a numerical study allows to select the dominant term: it is written as a $l$-series of eigenvalues, each one being characterized by a unique index $l$ whose origin is explained. Surprisingly $Z_N(0)$ shows a very simple exact closed-form expression *valid for any temperature*. The thermal study of the basic $l$-term allows to point out *crossovers* between $l$- and $(l+1)$-terms. Coming from high temperatures where the $l=0$-term is dominant and going to 0 K, $l$-eigenvalues showing increasing $l$-values are more and more selected. At $T = 0$ K $l \to +\infty$ and all the successive dominant $l$-eigenvalues become equivalent. As the $z$-spin correlation is null for $T > 0$ K but equal to 1 (in absolute value) for $T = 0$ K the critical temperature is $T_c = 0$ K. Using an analytical method similar to the one employed for $Z_N(0)$ we also give an exact expression *valid for any temperature* for the spin-spin correlations as well as for the correlation length $\xi$. In the $T=0$-limit we obtain a diagram of magnetic phases which is similar to the one derived through a renormalization approach. By taking the low-temperature limit of $\xi$ we obtain the same expressions as the corresponding ones derived through a renormalization process, for each zone of the magnetic phase diagram, thus bringing for the first time a strong validation to the full exact solution of the model *valid for any temperature*.

## 1. Introduction

Since the middle of the eighties with the discovery of high-temperature superconductors [1], the nonlinear σ-model has known a new interest for it allows to describe the properties of two-dimensional quantum antiferromagnets such as $La_2CuO_4$ [2-10]. These antiferromagnets, when properly doped, become superconductors up to a critical temperature $T_c$ notably high compared to other types of superconducting materials.

For studying the magnetic properties of such magnets Chakravarty *et al.* [6] have shown that it is necessary to consider the associated space-time which is composed of the crystallographic space of dimension $d$ to which a time-like axis, namely called the $i\tau$-axis, is added. The space-like axes are infinite but the time-like axis has a finite length called the "slab thickness" which is inversely proportional to the temperature $T$ and hence goes to infinity as $T$ goes to zero. As a result $D = d + 1$ is the space-time dimension and here $d = 2$ and $D = 2+1$. Thus the nonlinear σ-model in 2+1 dimensions has been conjectured to be equivalent at low temperatures to the two-dimensional Heisenberg model [11,12], which in turn can be derived from the Hubbard model in the large $U$-limit [13].

In a seminal paper, Chakravarty *et al.* [6] have studied this model using the method of one-loop renormalization group (RG) improved perturbation theory initially developed by Nelson and Pelkovits [14]. These authors have related the σ-model to the spin-1/2 Heisenberg model by simply considering [6,10]:

(i) a nearest-neighbour $s = 1/2$ antiferromagnetic Heisenberg Hamiltonian on a square lattice characterized by a large exchange energy;
(ii) very small interplanar couplings and spin anisotropies.

In addition, they have pointed out that the long-wavelength, low-energy properties are well described by a mapping to a *two-dimensional classical Heisenberg* magnet because all the effects of quantum fluctuations can be resorbed by means of adapted renormalizations of the coupling constants. A low-temperature diagram of magnetic phases has been derived. It is characterized by three different magnetic regimes: the *Renormalized Classical Regime* (RCR), the *Quantum Critical Regime* (QCR) and the *Quantum Disordered Regime* (QDR). For each of these regimes Chakravarty *et al.* [6] have given a closed-form expression of the correlation length ξ *exclusively valid near the critical point* $T_c = 0$ K. Finally these authors have shown that the associated critical exponent is $\nu = 1$.

A little bit later Hasenfratz and Niedermayer published a more detailed low-temperature expression of the correlation length for the Renormalized Classical Regime, exclusively [15]. Finally, also using a renormalization group technique, Chubukov *et al.* [10] reconsidered the work of Chakravarty *et al.* [6] by detailing the static but also the dynamic low-temperature magnetic properties of antiferromagnets described by a two-dimensional classical Heisenberg model. They notably published exact expressions of the correlation length ξ and the magnetic susceptibility χ (restricted to the case of compensated antiferromagnets), also *exclusively valid near the critical point* $T_c = 0$ K, for each of the three zones of the magnetic diagram.

From an experimental point of view, at the end of the nineties, the first two-dimensional magnetic compounds appeared [16-18]. Some of them were composed of sheets of classical spins (i.e., manganese ions of spin $S = 5/2$) well separated from each others by nonmagnetic organic ligands, thus ensuring very weak intersheet dipole-dipole interactions i.e., good two-dimensional magnetic properties. In other words the three-dimension magnetic ordering only appeared at temperatures $T_{3d}$ very close to absolute zero. The sheets of magnetic ions were characterized by a square unit cell. In addition, within each sheet, the ions were themselves largely separated by organic ligands, thus considerably diminishing the dipole-dipole interactions within the sheet and ensuring a quasi isotropic nature of coupling between first-nearest spin neighbours through a mechanism of superexchange [19]. *These two-dimensional magnetic compounds were the first ones whose low-temperature magnetic properties were characterized by a quantum critical regime.*

Therefore the necessity of fitting experimental susceptibilities as well as the important theoretical conclusions of the respective works of Chakravarty *et al.* [6] and Chubukov *et al.* [10] motivated us to focus again on the two-dimensional-classical O(3) model developed on a lattice showing a square unit cell and $(2N)^2$ sites [20-23].

The mathematical framework common to our first series of articles was the following one:

(i) we first considered the local exchange Hamiltonian $H_{i,j}^{ex}$ associated with each lattice site $(i,j)$ which is the carrier of a classical spin showing Heisenberg (isotropic) couplings with its first-nearest neighbours; in that case the evaluation of the zero-field partition function $Z_N(0)$ necessitates to expand each local operator $\exp(-\beta H_{i,j}^{ex})$ on the infinite basis of spherical harmonics $Y_{l,m}$;

(ii) each harmonics is thus characterized by a couple of integers $(l,m)$, with $l \geq 0$ and $m \in [-l, +l]$ and is nothing but the eigenfunction of each operator $\exp(-\beta H_{i,j}^{ex})$; the corresponding eigenvalue $\lambda_l(-\beta J)$ is the modified Bessel function of the first kind $(\pi/2\beta J)^{1/2} I_{l+1/2}(-\beta J)$ where $\beta = 1/k_B T$ is the Boltzmann factor and $J$ the exchange energy between consecutive spin neighbours.

As a result the polynomial expansion describing the zero-field partition function $Z_N(0)$ directly appears as a characteristic $l$-polynomial, for the considered lattice. Each term is composed of a radial factor i.e., a product of the $\lambda_{l_{i,j}}(-\beta J)$'s and an angular factor composed of a product of integrals $F_{i,j}$ containing spherical harmonics (with one eigenvalue $\lambda_{l_{i,j}}(-\beta J)$ and one eigenfunction i.e., one harmonics per bond).

We observed that, for most of the examined compounds showing a low-temperature *quantum critical regime*, when fitting the corresponding experimental susceptibilities, the characteristic $l$-polynomial associated with the theoretical susceptibility $\chi$ could be restricted to the dominant term characterized by $l = 0$, in the physical case of an infinite lattices. It also meant that the characteristic $l$-polynomial associated with $Z_N(0)$ was itself reduced to the term $l = 0$. In other words, for both characteristic $l$-polynomials which share a common part, no mathematical study was necessary in spite of the fact that this assumption gave good results for the involved exchange energies $J$ i.e., the exact corresponding tabulated experimental values, with a Landé factor value very close to the theoretical one $G = 2$ (in $\mu_B/\hbar$ unit). However we also discovered that, for some compounds characterized by the same low-temperature quantum critical regime, it was necessary to take into account the terms $l = 0$ but also $l = 1$ (with $m = 0$) in the $l$-expansion of $\chi$ for obtaining a good fit of experimental susceptibilities.

Thus, from a theoretical point of view, the condition leading to choose the term $l = 0$ exclusively or the terms $l = 0$ and $l = 1$ ($m = 0$) in the common $l$-polynomial part shared by $Z_N(0)$ and $\chi$ remained a puzzling question. For all the experimental fits, the lowest possible value reached by temperature for ensuring a pure two-dimensional magnetic behaviour was $T = T_{3d}$ when the 3d-magnetic ordering appears. We then observed that, if restricting the $l$-expansion of $\chi$ to the term $l = 0$, we had to fulfil the numerical condition $k_B T_{3d}/|J| \geq 0.255$. But, if compelled to consider the terms $l = 0$ and $l = 1$ ($m = 0$), we had $k_B T_{3d}/|J| \geq 0.043$. Finally the low-temperature theoretical diagram of magnetic phases was restricted to a single phase, the *quantum critical regime*, in contradiction with the results derived near $T_c = 0$ K, from a renormalization technique which points out three different magnetic regimes [6,10].

In order to solve these difficulties a full study of the characteristic $l$-polynomial associated with $Z_N(0)$ appeared as unavoidable. This is the aim of the present paper. Even if starting with the same mathematical considerations common to the first series of papers previously published [20-23] this paper is intended as a new work because, in section 2 and for the first time, we establish the complete closed-form expression of the characteristic $l$-polynomial associated with $Z_N(0)$, *valid for any lattice size, any temperature and any l*.

The examination of the case of a finite lattice is out of the framework of the present article [24]. Then we exclusively consider the physical case of an infinite lattice (i.e., the thermodynamic limit) in section 2. We numerically show that, if studying the angular part of each $l$-term of the characteristic $l$-polynomial, the value $m = 0$ is selected. In addition we formally prove that the higher-degree term of the characteristic $l$-polynomial giving $Z_N(0)$ is such as all the $l$'s are equal to a common value $l_0$. Surprisingly we then obtain a very simple closed-form expression for $Z_N(0)$, *valid for any temperature and any l*.

Finally, in section 2, we report a further thermal numerical study of the $l$-higher-degree term. Thus and for the first time, this study allows to point out a *new result* i.e., *thermal crossovers* between two consecutive $l$- and ($l+1$)-eigenvalues. It means that the characteristic $l$-polynomial can be reduced to a single $l$-term within a given temperature range but, for the whole temperature range, all the $l$-eigenvalues must be kept.

Thus, if coming from high temperatures, we now explain why the value $l = 0$ characterizes the dominant term for reduced temperatures such as $k_B T/|J| \geq 0.255$. For $0.255 \geq$ $k_B T/|J| \geq 0.043$ we have $l = 1$ and so on. Finally $l$-eigenvalues $\lambda_l(\beta|J|)$, with increasing $l > 0$, are successively dominant when temperature is decreasing down to 0 K. In the vicinity of absolute zero the dominant term is characterized by $l \to +\infty$. As all the $l$-eigenvalues show a very close low-temperature behaviour we then deal with a continuous spectrum of eigenvalues, confirming the fact that the critical temperature is $T_c = 0$ K, in agreement with Mermin-Wagner's theorem [25].

In section 3 we analytically show that the spin correlation is such as $<S^z> = 0$ for $T > 0$ K whereas $<S^z> = \pm 1$ for $T = 0$ K, again confirming the fact that $T_c = 0$ K. Then, for the first time, in the thermodynamic limit, we obtain *the exact closed-form expression of the spin-spin correlation* $<S_{0,0}.S_{k,k'}>$ *between any couple of lattice sites* (0,0) *and* (k,k'), *valid for any temperature*.

*For doing so we first show that all the correlation paths are confined within a closed domain called the "correlation domain" which is a rectangle whose sides are the bonds linking sites* (0,0), (0,k'), (k,k') *and* (k,0) *(theorem 1). Second we prove that open or closed loops are forbidden so that all the correlation paths show the same shortest possible length between any couple of lattice sites. All of them have the same weight i.e., they are composed of the same number of horizontal (respectively, vertical) bonds as the horizontal (respectively, vertical) sides of the correlation domain (theorem 2).* This allows to derive *an exact expression of the correlation length* ξ *also valid for any temperature*.

In section 4 we examine the low-temperature behaviour of the $\lambda_l(\beta|J|)$'s. We retrieve the low-temperature magnetic phase diagram. It is strictly similar to the one derived from a renormalization technique [6,10]. The low-temperature magnetic properties are described in terms of universal parameters $k_B T/\rho_s$ and $k_B T/\Delta$ where $\rho_s$ and $\Delta$ are the spin stiffness and the $T=0$-energy gap between the ground state and the first exited one, respectively. By taking the low-temperature limit of the correlation length ξ we obtain the same expressions as the corresponding ones derived through a renormalization process, for each zone of the magnetic phase diagram, thus bringing for the first time a strong validation to the full exact solution of the model *valid for any temperature*. At $T_c = 0$ K we retrieve the critical exponent $\nu = 1$, as previously shown [6,10].

In addition, near the critical point, the correlation length $\xi_x$ can be simply expressed owing to the absolute value of the renormalized spin-spin correlation $|<\widetilde{S_{0,0}.S_{0,1}}>|$ between first-nearest neighbours i.e., sites (0,0) and (0,1). In addition $<\widetilde{S_{0,0}.S_{0,1}}>$ and $\xi_x$ can also be written with the derivative of the logarithm of the dominant eigenvalue $\lambda_l(\beta|J|)$ with respect to $\beta|J|$, in the limit $l \to +\infty$, thus justifying the detailed study of $Z_N(0)$ in this article.

Section 5 summarizes our conclusions.

The appendix gives all the detailed calculations necessary for understanding the main text, notably the low-temperature study of key physical parameters.

## 2. Exact expression of the zero-field partition function of an infinite lattice

### 2.1 Definitions

The general Hamiltonian describing a lattice characterized by a square unit cell composed of $(2N + 1)^2$ sites, each one being the carrier of a classical spin $S_{i,j}$, is given by:

$$H = \sum_{i=-N}^{N} \sum_{j=-N}^{N} (H_{i,j}^{ex} + H_{i,j}^{mag}), \quad (1)$$

with $N \to +\infty$ in the case of an infinite lattice on which we exclusively focus in this article and

$$H_{i,j}^{ex} = (J_1 S_{i,j+1} + J_2 S_{i+1,j}).S_{i,j}, \quad (2)$$

$$H_{i,j}^{\mathrm{mag}} = -G_{i,j} S_{i,j}^{z} B, \tag{3}$$

where:

$$G_{i,j} = G \text{ if } i + j \text{ is even or null}, \quad G_{i,j} = G' \text{ if } i + j \text{ is odd}. \tag{4}$$

In equation (2) $J_1$ and $J_2$ refer to the exchange interaction between first-nearest neighbours belonging to the horizontal lines and vertical rows of the lattice, respectively. $J_i > 0$ (respectively, $J_i < 0$, with $i = 1, 2$) denotes an antiferromagnetic (respectively, ferromagnetic) coupling. $G_{i,j}$ is the Landé factor characterizing each spin $S_{i,j}$ and expressed in $\mu_B/\hbar$ unit. Finally we consider that the classical spins $S_{i,j}$ are unit vectors so that the exchange energy $JS(S + 1) \sim JS^2$ is written $J$. It means that we do not take into account the number of spin components in the normalization of $S_{i,j}$'s so that $S^2 = 1$.

When $H_{i,j}^{\mathrm{mag}} = 0$ the zero-field partition function $Z_N(0)$ is defined as:

$$Z_N(0) = \prod_{i=-N}^{N} \prod_{j=-N}^{N} \int dS_{i,j} \exp\left(-\beta \sum_{i=-N}^{N} \sum_{j=-N}^{N} H_{i,j}^{\mathrm{ex}}\right), \tag{5}$$

where $\beta = 1/k_B T$ is the Boltzmann factor. In other words the zero-field partition function $Z_N(0)$ is simply obtained by integrating the operator $\exp(-\beta H^{\mathrm{ex}})$ over all the angular variables characterizing the states of all the classical spins belonging to the lattice.

## 2.2 Preliminaries

Due to the presence of classical spin moments, all the operators $H_{i,j}^{\mathrm{ex}}$ commute and the exponential factor appearing in the integrand of equation (5) can be written:

$$\exp\left(-\beta \sum_{i=-N}^{N} \sum_{j=-N}^{N} H_{i,j}^{\mathrm{ex}}\right) = \prod_{i=-N}^{N} \prod_{j=-N}^{N} \exp\left(-\beta H_{i,j}^{\mathrm{ex}}\right). \tag{6}$$

As a result, the particular nature of $H_{i,j}^{\mathrm{ex}}$ given by equation (2) allows one to separate the contributions corresponding to the exchange coupling involving classical spins belonging to the same horizontal line $i$ of the layer (i.e., $S_{i,j-1}$, $S_{i,j+1}$ and $S_{i,j}$) or to the same vertical row $j$ (i.e., $S_{i-1,j}$, $S_{i+1,j}$ and $S_{i,j}$). In fact, for each of the four contributions (one per bond connected to the site $(i,j)$ carrying the spin $S_{i,j}$), we have to expand a term such as $\exp(-AS_1 \cdot S_2)$ where $A$ is $\beta J_1$ or $\beta J_2$ (the classical spins $S_1$ and $S_2$ being considered as unit vectors). If we call $\Theta_{1,2}$ the angle between vectors $S_1$ and $S_2$, characterized by the couples of angular variables $(\theta_1, \varphi_1)$ and $(\theta_2, \varphi_2)$, it is possible to expand the operator $\exp(-A\cos\Theta_{1,2})$ on the infinite basis of spherical harmonics which are eigenfunctions of the angular part of the Laplacian operator on the sphere of unit radius $S^2$:

$$\exp(-A\cos\Theta_{1,2}) = 4\pi \sum_{l=0}^{+\infty} \left(\frac{\pi}{2A}\right)^{1/2} I_{l+1/2}(-A) \sum_{m=-l}^{+l} Y_{l,m}^*(S_1) Y_{l,m}(S_2). \tag{7}$$

In the previous equation the $(\pi/2A)^{1/2} I_{l+1/2}(-A)$'s are modified Bessel functions of the first kind; $S_1$ and $S_2$ symbolically represent the couples $(\theta_1, \varphi_1)$ and $(\theta_2, \varphi_2)$. If we set:

$$\lambda_l(-\beta j) = \left(\frac{\pi}{2\beta j}\right)^{1/2} I_{l+1/2}(-\beta j), \quad j = J_1 \text{ or } J_2, \tag{8}$$

each operator $\exp\left(-\beta H_{i,j}^{\mathrm{ex}}\right)$ is finally expanded on the infinite basis of eigenfunctions (the spherical harmonics), whereas the $\lambda_l$'s are nothing but the associated eigenvalues. Under these conditions, the zero-field partition function $Z_N(0)$ directly appears as a characteristic polynomial.

In the case of an infinite lattice edge effects are negligible so that it is equivalent to consider a lattice wrapped on a torus characterized by two infinite radii of curvature. Ho-rizontal lines $i = -N$ and $i = N$ on the one hand and vertical lines $j = -N$ and $j = N$ on the other one are confused so that there are $(2N)^2$ sites and $2(2N)^2$ bonds, with $N \to +\infty$. As a result $Z_N(0)$ can be written as:

$$Z_N(0) = (4\pi)^{8N^2} \prod_{i=-(N-1)}^{N} \prod_{j=-(N-1)}^{N} \sum_{l_{i,j}=0}^{+\infty} \sum_{l'_{i,j}=0}^{+\infty} \sum_{m_{i,j}=-l_{i,j}}^{+l_{i,j}} \sum_{m'_{i,j}=-l'_{i,j}}^{+l'_{i,j}} F_{i,j} \lambda_{l_{i,j}}(-\beta J_1) \lambda_{l'_{i,j}}(-\beta J_2), \quad (9)$$

$$F_{i,j} = \int dS_{i,j} Y_{l'_{i+1,j},m'_{i+1,j}}(S_{i,j}) Y_{l_{i,j-1},m_{i,j-1}}(S_{i,j}) Y^*_{l_{i,j},m_{i,j}}(S_{i,j}) Y^*_{l'_{i,j},m'_{i,j}}(S_{i,j}) \quad (10)$$

where $F_{i,j}$ is the current integral per site (with one spherical harmonics per bond).

Using the following decomposition of any product of two spherical harmonics appearing in the integrand of $F_{i,j}$ [26]

$$Y_{l_1,m_1}(S) Y_{l_2,m_2}(S) = \sum_{L=|l_1-l_2|}^{l_1+l_2} \sum_{M=-L}^{+L} \left[ \frac{(2l_1+1)(2l_2+1)}{4\pi(2L+1)} \right]^{1/2} C_{l_1 \, 0 \, l_2 \, 0}^{L \, 0} C_{l_1 \, m_1 \, l_2 \, m_2}^{L \, M} Y_{L,M}(S) \quad (11)$$

where $C_{l_1 \, m_1 \, l_2 \, m_2}^{l_3 \, m_3}$ is a Clebsch-Gordan (C.G.) coefficient and the orthogonality relation of spherical harmonics $F_{i,j}$ can be expressed as the following C.G. series

$$F_{i,j} = \frac{1}{4\pi} \left[ (2l'_{i+1,j}+1)(2l_{i,j-1}+1)(2l_{i,j}+1)(2l'_{i,j}+1) \right]^{1/2} \sum_{L_{i,j}=L_<}^{L_>} \frac{1}{2L_{i,j}+1} \times$$

$$\times \sum_{M_{i,j}=-L_{i,j}}^{+L_{i,j}} C_{l'_{i+1,j} \, 0 \, l_{i,j-1} \, 0}^{L_{i,j} \, 0} C_{l'_{i+1,j} \, m'_{i+1,j} \, l_{i,j-1} \, m_{i,j-1}}^{L_{i,j} \, M_{i,j}} C_{l_{i,j} \, 0 \, l'_{i,j} \, 0}^{L_{i,j} \, 0} C_{l_{i,j} \, m_{i,j} \, l'_{i,j} \, m'_{i,j}}^{L_{i,j} \, M_{i,j}}. \quad (12)$$

The C.G. coefficients $C_{l_{i,j} \, m_{i,j} \, l'_{i,j} \, m'_{i,j}}^{L_{i,j} \, M_{i,j}}$ and $C_{l'_{i+1,j} \, m'_{i+1,j} \, l_{i,j-1} \, m_{i,j-1}}^{L_{i,j} \, M_{i,j}}$ (with $M_{i,j} \neq 0$ or $M_{i,j}$ = 0) do not vanish if the triangular inequalities $|l_{i,j} - l'_{i,j}| \leq L_{i,j} \leq l_{i,j} + l'_{i,j}$ and $|l'_{i+1,j} - l_{i,j-1}| \leq L_{i,j} \leq l'_{i+1,j} + l_{i,j-1}$ are fulfilled, respectively. As a result, we must have $L_< = \max(|l'_{i+1,j} - l_{i,j-1}|, |l_{i,j} - l'_{i,j}|)$ and $L_> = \min(l'_{i+1,j} + l_{i,j-1}, l_{i,j} + l'_{i,j})$.

## 2.3 Principles of construction of the characteristic polynomial associated with the zero-field partition function

The zero-field partition function given by equation (9) can be rewritten under the general form

$$Z_N(0) = (4\pi)^{8N^2} \prod_{i=-(N-1)}^{N} \prod_{j=-(N-1)}^{N} \sum_{l_{i,j}=0}^{+\infty} \sum_{l'_{i,j}=0}^{+\infty} \sum_{m_{i,j}=-l_{i,j}}^{+l_{i,j}} \sum_{m'_{i,j}=-l'_{i,j}}^{+l'_{i,j}} u_{l_{i,j},l'_{i,j}}(T) \quad (13)$$

with:

$$u_{l_{i,j},l'_{i,j}}(T) = F_{i,j} \lambda_{l_{i,j}}(-\beta J_1) \lambda_{l'_{i,j}}(-\beta J_2). \quad (14)$$

The examination of equation (13) giving the polynomial expansion of $Z_N(0)$ allows one to say that its writing is nothing but that one derived from the formalism of the transfer-matrix technique. Each current term appears as a product of two subterms:

(i) a *temperature-dependent radial factor* containing a product of the various eigenvalues $\lambda_l(-\beta j)$, $j = J_1$ or $J_2$, of the full lattice operator $\exp(-\beta H^{ex})$ (with one eigenvalue per bond);

(ii) an *angular factor* containing a product of integrals $F_{i,j}$ composed of spherical harmonics (the eigenfunctions) describing all the spin states of all the lattice sites (with one integral per site).

Equation (13) can also be artificially shared into two parts labelled Part I and Part II of respective zero-field partition functions $Z_N^I(0)$ and $Z_N^{II}(0)$ so that the zero-field partition function $Z_N(0)$ can be written as

$$Z_N(0) = Z_N^I(0) + Z_N^{II}(0), \quad (15)$$

with

$$Z_N^I(0) = (4\pi)^{8N^2} \sum_{l=0}^{+\infty} \prod_{i=-(N-1)}^{N} \prod_{j=-(N-1)}^{N} \sum_{m_{i,j}=-l}^{+l} \sum_{m'_{i,j}=-l}^{+l} u_{l,l}(T)^{4N^2},$$

$$Z_N^{II}(0) = (4\pi)^{8N^2} \prod_{i=-(N-1)}^{N} \prod_{j=-(N-1)}^{N} \sum_{l_{i,j}=0}^{+\infty} \sum_{\substack{l'_{i,j}=0,\\l_{i,j}\neq l'_{i,j}}}^{+\infty} \sum_{m_{i,j}=-l_{i,j}}^{+l_{i,j}} \sum_{m'_{i,j}=-l'_{i,j}}^{+l'_{i,j}} u_{l_i,l_j}(T) \quad (16)$$

where $u_{l_i,l_j}(T)$ is given by equation (14). As a result Part I contains the general term $\left[F_{i,j}\lambda_l(-\beta J_1)\lambda_l(-\beta J_2)\right]^{(2N)^2}$ i.e., all the bonds are characterized by the same integer $l$ but we can have a set of different relative integers $m_{i,j} \in [-l, +l]$ and $m'_{i,j} \in [-l, +l]$ with $m_{i,j} = m'_{i,j}$ or $m_{i,j} \neq m'_{i,j}$. Part II appears as a product of "cluster" terms such as $\left[F_{i,j}\lambda_{l_k}(-\beta J_1)\lambda_{l'_k}(-\beta J_2)\right]^{n_k}$ with $n_k < (2N)^2$ and the condition $n_1 + n_2 + ... + n_k = (2N)^2$. Thus, only $n_k$ bonds are characterized by the same integers $l_k$, $l'_k$ and a collection of different relative integers $m_{i,j} \in [-l_k, +l_k]$ and $m'_{i,j} \in [-l'_k, +l'_k]$, with $m_{i,j} = m'_{i,j}$ or $m_{i,j} \neq m'_{i,j}$.

## 2.4 General selection rules for the whole lattice

The non-vanishing condition of each current integral $F_{i,j}$ due to that of C.G. coefficients allows one to derive two types of universal *selection rules* which are *temperature-independent*.

The *first selection rule* concerns the coefficients $m$ and $m'$ appearing in equation (12). We have $(2N)^2$ equations (one per lattice site) such as:

$$m_{i,j-1} + m'_{i+1,j} - m_{i,j} - m'_{i,j} = 0. \quad (SRm) \quad (17)$$

At this step we must note that, if each spherical harmonics $Y_{l,m}(S) = Y_{l,m}(\theta,\varphi)$ appearing in the integrand of $F_{i,j}$ is replaced by its own definition i.e., $C_l^m \exp(im\varphi)P_l^m(\cos\theta)$ where $C_l^m$ is a constant depending on coefficients $l$ and $m$ [26] and $P_l^m(\cos\theta)$ is the associated Legendre polynomial, the non-vanishing condition of the $\varphi$-part directly leads to equation (17). As a result, we can make two remarks: the SRm relation is unique; due to the fact that the $\varphi$-part of the $F_{i,j}$-integrand is null, $F_{i,j}$ is a pure real number.

The *second selection rule* is derived from the fact that the various coefficients $l$ and $l'$ appearing in equation (12) obey triangular inequalities as noted after this equation [24]. If $M_{i,j} \neq 0$ the determination of $l_{i,j}$'s and $l'_{i,j}$'s is exclusively numerical. If $M_{i,j} = 0$ we have a more restrictive vanishing condition [26]:

$$C_{l_1\ 0\ l_2\ 0}^{l_3\ 0} = 0, \text{ if } l_1 + l_2 + l_3 = 2g+1,$$

$$C_{l_1\ 0\ l_2\ 0}^{l_3\ 0} = (-1)^{g-l_3}\sqrt{2l_3+1}\ K, \text{ if } l_1 + l_2 + l_3 = 2g, \quad (18)$$

where $K$ is a coefficient depending on $l_1$, $l_2$, $l_3$ and $g$ [26]. In equation (12) $C_{l_{i,j}\ 0\ l'_{i,j}\ 0}^{L_{i,j}\ 0}$ does not vanish if $l_{i,j} + l'_{i,j} + L_{i,j} = 2A_{i,j} \geq 0$ whereas, for $C_{l'_{i+1,j}\ 0\ l_{i,j-1}\ 0}^{L_{i,j}\ 0}$, we must have $l_{i,j-1} + l'_{i+1,j} + L_{i,j} = 2A'_{i,j} \geq 0$. Thus, if summing or substracting the two previous equations over $l$ and $l'$, we have $(2N)^2$ equations (one per lattice site) such as:

$$l_{i,j-1} + l'_{i+1,j} + l_{i,j} + l'_{i,j} = 2g_{i,j}, \quad (SRl1)$$

$$l_{i,j-1} + l'_{i+1,j} - l_{i,j} - l'_{i,j} = 2g'_{i,j}, \quad (SRl2) \tag{19}$$

(or equivalently $l_{i,j} + l'_{i,j} - l_{i,j-1} - l'_{i+1,j} = 2g''_{i,j}$, with $g''_{i,j} = -g'_{i,j}$) where $g'_{i,j}$ or $g''_{i,j}$ is a relative integer. We obtain two types of equations which are similar to equation (17) but now, in equation (19), instead of having a null right member like in equation (17), we can have a positive, null or negative but always even second member.

## 2.5 Zero-field partition function in the thermodynamic limit

The case of thermodynamic limit ($N \to +\infty$) is less restrictive than that of a finite lattice studied in previous articles [24] because we deal with a purely numerical problem. For simplifying the discussion when necessary we shall restrict the study to the case $J = J_1 = J_2$ without loss of generality. If not specified we shall refer to the general case $J_1 \neq J_2$.

As previously explained in Subsec. 2.3 the characteristic polynomial associated with $Z_N(0)$ for $N$ finite or infinite is composed of two parts: Part I of current term $\left[F_{i,j}\lambda_l(-\beta J_1)\lambda_l(-\beta J_2)\right]^{(2N)^2}$ and Part II whose current term is a product of "cluster" terms $\left[F_{i,j}\lambda_l(-\beta J_1)\lambda_{l'}(-\beta J_2)\right]^{n_k}$, with $n_k < (2N)^2$ and $\sum_k n_k = (2N)^2$, so that, in both cases, the numerical study concerns the common term $F_{i,j}\lambda_l(-\beta J_1)\lambda_{l'}(-\beta J_2)$ with $l = l'$ or $l \neq l'$. The $m$'s and $l$'s (respectively the $m'$'s and $l'$'s) appearing in $F_{i,j}$ (cf equation (12)) can vary or not from one site to another site.

(i) First let us consider the case of the $m$'s. We have to solve a linear system of $(2N)^2$ equations (17) (one per site) but with $2(2N)^2$ unknowns $m_{i,j}$ and $m'_{i,j}$. As it remains $2(2N)^2 - (2N)^2 = (2N)^2$ independent solutions over the set $\mathbb{Z}$ of relative integers $m_{i,j}$ and $m'_{i,j}$ (with here $N \to +\infty$) *it means that there are $(2N)^2$ different expressions (i.e., here an infinity) for each local angular factor appearing in each term of the characteristic polynomial giving $Z_N(0)$ so that the statistical problem remains unsolved.* Thus, at first sight, this result means that *there is no unique expression for $Z_N(0)$.*

In summary there are only $(2N)^2$ independent solutions i.e., $(2N)^2$ different sets of coefficients $(m_{i,j}, m'_{i,j})$ obeying equation (17). The simplest solution is given by the condition $|m_{i,j}| = |m'_{i,j}| \neq 0$ or $m_{i,j} = m'_{i,j} = 0$, for the whole lattice.

(ii) The study of $l$'s is strictly similar to that of $m$'s because we have to solve a linear system of $(2N)^2$ equations (19) (one per site) with $2(2N)^2$ unknowns $l_{i,j}$ and $l'_{i,j}$. We have $(2N)^2$ independent solutions over the set $\mathbb{N}$ of integers $l_{i,j}$ and $l'_{i,j}$ and the particular solution $l_{i,j} = l'_{i,j} \neq 0$ or $l_{i,j} = l'_{i,j} = 0$, for the whole lattice.

In other words, when $N \to +\infty$, a separate numerical study of integrals $F_{i,j}$ must allow one to select a unique $m$-value so that $F_{i,j}$ is maximum. First we restrict the set of integers $l_{i,j}$, $l_{i-1,j}$, $l'_{i,j}$ and $l'_{i+1,j}$ appearing in the integrand of $F_{i,j}$ to two different integers $l_i$ and $l_j$. In that case, if setting $F_{i,j} = F(l_i, l_j, m)$ with $l_i = l_j \geq 0$ or $l_i \geq l_j$, we have

$$F_{i,j} = F(l_i, l_j, m) = \frac{(2l_i+1)(2l_j+1)}{4\pi} \sum_{L=|l_i-l_j|}^{l_i+l_j} \frac{1}{2L+1}\left[C_{l_i\ 0\ l_j\ 0}^{L\ 0} C_{l_i\ m\ l_j\ m}^{L\ 2m}\right]^2. \tag{20}$$

In the infinite-lattice limit we expect that the highest eigenvalue must naturally arise in Part I of current term $\left[F_{i,j}\lambda_l(-\beta J_1)\lambda_l(-\beta J_2)\right]^{(2N)^2}$. If using equation (14) with $l_{i,j} = l'_{i,j} = l$, let $u_{\max} = F_{i,j}\lambda_L(-\beta J_1)\lambda_L(-\beta J_2)$ (where $L = l_{\max}$) be this contribution. We make the assumption that it dominates all the other terms inside Part I as well as all the ones composing Part II defined by equations (15) and (16) [24]. This can occur in the whole temperature range or in a smaller temperature one if there exist thermal crossover phenomena among the set of eigenvalues.

In that case the dominant eigenvalue $\lambda_l(-\beta J_i) = \lambda_l(-J_i/k_B T)$ over a temperature range becomes subdominant when the temperature $T$ is outside this range and a new eigenvalue previously subdominant becomes dominant and so on. In the case of 1$d$ spin chains, we always have the same highest eigenvalue (within the framework described previously).

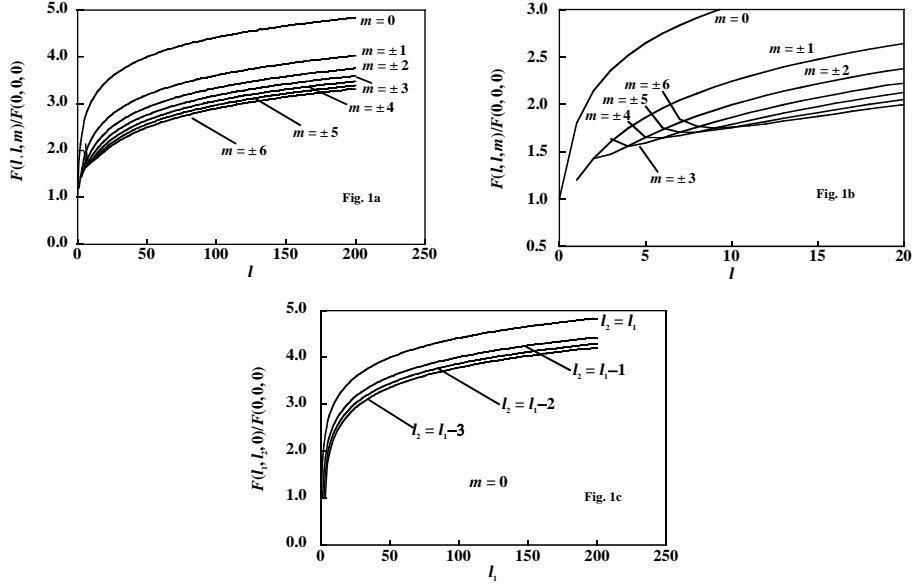

**Figure 1.** a) Numerical study of the ratio $F(l, l, m)/F(0, 0, 0)$ *vs* $l$ for various values of $m$ ($F(l, l, m)$ is given by equation (20) and $F(0, 0, 0) = 1/4\pi$); b) zoom for reduced $l$-values; c) numerical study of the ratio $F(l_1, l_2, 0)/F(0, 0, 0)$ for various values of $l_2 \leq l_1$.

In this respect we have first studied the integral $F_{i,j} = F(l_i, l_j, m)$ given by equation (20) with $l_i = l_j = l \geq 0$. In Fig. 1a we have reported the ratio $F(l,l,m)/F(0,0,0)$ *vs* $l$ for various $m$-values such as $|m| \leq l$ (with $F(0,0,0) = 1/4\pi$). We immediately observe that this ratio rapidly decreases for increasing $|m|$-values, for any $l$. However, we have zoomed the beginning of each curve corresponding to the case $|m| = l$. This trend is not followed but we always have $F(l,l,m) < F(l,l,0)$ (see Fig. 1b). Second, in Fig. 1c, for $m = 0$, we observe that $F(l_i, l_j, 0)$ decreases for $l_j < l_i$. As a result, when $N \to +\infty$, the integral $F(l,l,0)$ obtained if $l = l_i = l_j$ appears as the dominant one i.e.,

$$[F(l,l,0)]^{(2N)^2} \gg [F(l,l,m)]^{(2N)^2} \gg [F(l_i,l_j,m)]^{(2N)^2}, \ l_i = l \geq l_j, \ \text{as } N \to +\infty, \qquad (21)$$

so that the value $m = 0$ is selected. In addition this result shows that it is not necessary to consider 4 different integers $l_{i,j}$ and $l'_{i,j}$ in the integral $F(l_i, l_j, m)$.

For sake of simplicity we now restrict to the case $J = J_1 = J_2$. Under these conditions equation (13) can be rewritten in the thermodynamic limit:

$$Z_N(0) = (4\pi)^{8N^2} \left[ \sum_{l=0}^{+\infty} \left[ F(l,l,0) \lambda_l (-\beta J)^2 \right]^{4N^2} \right.$$

$$\left. + \prod_{i=-(N-1)}^{N} \prod_{j=-(N-1)}^{N} {\sum_{l_i=0}^{+\infty}}' {\sum_{l_j=0}^{+\infty}}' F(l_i, l_j, 0) \lambda_{l_i}(-\beta J) \lambda_{l_j}(-\beta J) \right], \text{ as } N \to +\infty. \quad (22)$$

The notation ${\sum_{l_i=0}^{+\infty}}' {\sum_{l_j=0}^{+\infty}}'$ means that $l_i$ and $l_j$ are chosen so that the corresponding current second-rank term cannot give back the first-rank one in which $l_i = l_j = l$.

In a first step we must wonder if all the current terms of the previous $l$-series must be kept in the first term of equation (22) i.e., if the series must be truncated, for a given range of temperature $[T_{l_i,<}, T_{l_i,>}]$. As a result, for any $T \in [T_{l_i,<}, T_{l_i,>}]$, we define the dominant term

$$u_{\max} = f_{L,L} \lambda_L (-\beta J)^2, \ f_{L,L} = F(L,L,0), \ L = l_{\max} \qquad (23)$$

where $f_{L,L} = F(L,L,0)$ is given by equation (10) reduced to $m = m' = 0$ as well as the following ratio:

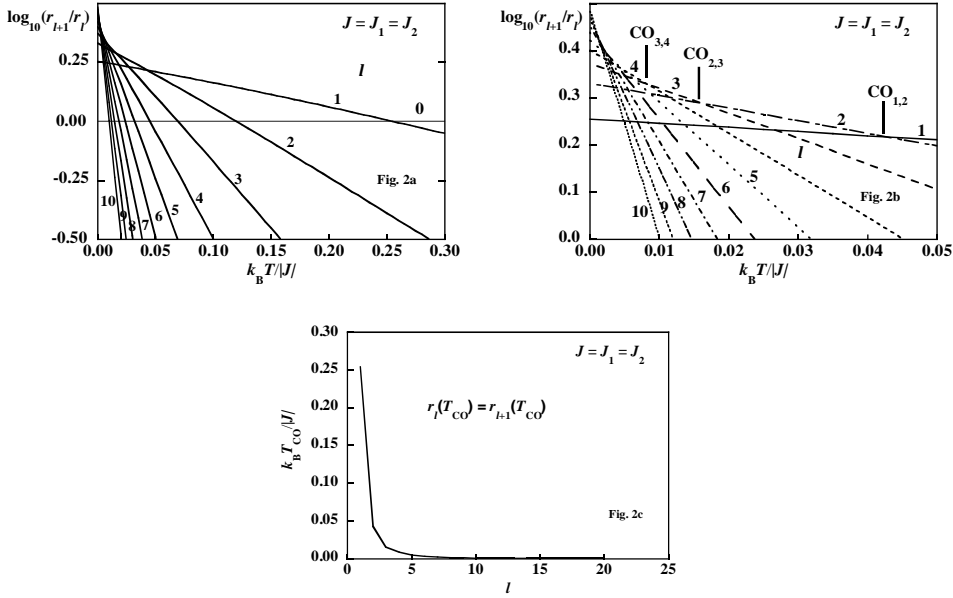

**Figure 2.** a) Thermal variations of $\log_{10}(r_{l+1}/r_l)$ for various values of $l$ where the ratio $r_{l+1}/r_l$ is defined by equation (24); b) zoom of the plot allowing to have a better insight of the crossover phenomena between various $l$-regimes; c) plot of the crossover temperature $T_{CO}$ vs $l$.

$$\frac{r_{l+1}}{r_l} = \frac{f_{l+1,l+1}}{f_{l,l}} \left( \frac{\lambda_{l+1}(-\beta J)}{\lambda_l(-\beta J)} \right)^2, \quad J = J_1 = J_2. \tag{24}$$

We have studied the thermal behaviour of $r_{l+1}/r_l$ for various finite $l$-values. This work is reported in Fig. 2a. We observe that $\log_{10}(r_{l+1}/r_l)$ shows a decreasing linear behaviour with respect to $k_B T/|J|$. We have zoomed Fig. 2a in the very low-temperature domain

(Fig. 2b). If $r_{l+1}/r_l < 1$ $\log_{10}(r_{l+1}/r_l) < 0$ and $\log_{10}(r_{l+1}/r_l) > 0$ if $r_{l+1}/r_l > 1$. We can then point out a succession of *crossovers*, each crossover being characterized by a specific temperature called *crossover temperature* $T_{CO}$. $T_{CO}$ is the solution of the equation:

$$r_l(T_{CO}) = r_{l+1}(T_{CO}) \tag{25}$$

i.e., owing to equation (24):

$$\frac{\lambda_{l+1}(|J|/k_B T_{CO})}{\lambda_l(|J|/k_B T_{CO})} = \left[ \frac{f_{l,l}}{f_{l+1,l+1}} \right]^{1/2}. \tag{26}$$

For instance, for the reduced temperatures such as $k_B T/|J| \geq 0.255$, the value $l = 0$ is dominant i.e., $\lambda_0(-\beta J)$ represents the dominant term of the characteristic polynomial. All the other terms $\lambda_l(-\beta J)$ with $l > 0$ are subdominant. When $0.255 \geq k_B T/|J| \geq 0.043$ $l = 1$ is dominant so that $\lambda_1(-\beta J)$ is now the dominant term of the characteristic polynomial whereas $\lambda_0(-\beta J)$ has become the subdominant one as well as all the other terms $\lambda_l(-\beta J)$ with $l \neq 1$ etc... In that case the crossover temperature corresponding to the transition between the regimes respectively characterized by $l = 0$ and $l = 1$ is labelled $T_{CO_{0,1}}$. We have reported $k_B T_{CO}/|J|$ vs $l$ in Fig. 2c.

As expected we observe that $T_{CO}$ rapidly decreases when $l$ increases. It means that, when the temperature tends to absolute zero, it appears a succession of closer and closer crossovers so that all the $l$-eigenvalues, characterized by an increasing $l$-value, successively play a role. But, when $T \approx 0$ K, all these eigenvalues intervene due to the fact that the crossover temperatures are closer and closer. *The discret eigenvalue spectrum tends* *to a continuum*. As a result we can say that $T = 0$ K *plays the role of critical temperature $T_c$*. This aspect will be more detailed later.

How interpreting this phenomena? In the $1d$-case (infinite spin chain) we always have $\lambda_0(-\beta J)$ as dominant eigenvalue in the whole range of temperature, the integral $F_{0,0} = f_{0,0}$ being always equal to unity. In the $2d$-case the situation is more complicated. The appearance of successive predominant eigenvalues is due to the presence of integrals $f_{l,l} \neq 1$, for any $l > 0$. A numerical fit shows that the ratio $f_{l,l}/f_{0,0}$ increases with $l$ *according to a logarithmic law*, more rapidly than the ratio $|\lambda_l(-\beta J)/\lambda_0(-\beta J)|^2$ which decreases when $l$ increases for a given temperature. The particular case $l \to +\infty$ will be examined in a forthcoming section.

Now, if taking into account the previous study, equation (22) can be rewritten as:

$$Z_N(0) = (4\pi)^{8N^2} [u_{\max}(T)]^{4N^2} \{1 + S_1(N,T) + S_2(N,T)\}, \quad T \in [T_{l_i,<}, T_{l_i,>}] \quad (27)$$

where $u_{\max}$ is given by equation (23) with $l_{\max} = l_i = l_j = l$ and:

$$S_1(N,T) = \sum_{l=0, l \neq l_{\max}}^{+\infty} \left[\frac{u_{l,l}(T)}{u_{\max}(T)}\right]^{4N^2}, \quad S_2(N,T) = \prod_{i=-(N-1)}^{N} \prod_{j=-(N-1)}^{N} \sum_{l_i=0}^{+\infty} \sum_{\substack{l_j=0, \\ l_j \neq l_i}}^{+\infty} \frac{u_{l_i,l_j}(T)}{u_{\max}(T)}. \quad (28)$$

As the ratios $|u_{l_i,l_i}(T)/u_{\max}|$ and $|u_{l_i,l_j}(T)/u_{\max}|$, $l_i \neq l_j$ (where $u_{l_i,l_i}(T)$ and $u_{l_i,l_j}(T)$ are given by equation (14)) are positive and lower than unity $S_1(N,T)$ and $S_2(N,T)$ are *absolutely convergent series*.

In Appendix *A.1* we have studied $Z_N(0)$ in the thermodynamic limit ($N \to +\infty$), for temperatures $T > 0$ K, in the whole range $[0+\varepsilon, +\infty[$, with $\varepsilon \ll 1$. We show that, for a given range $[T_{l_i,<}, T_{l_i,>}]$, $S_1(N,T) \gg S_2(N,T)$ (*cf* equation (A.6)) so that for any $T$

(i) $1 + S_1(N,T) \gg S_2(N,T)$ i.e., owing to equations (15) and (16) $Z_N^I(0) \gg Z_N^{II}(0)$ and $Z_N(0) \sim Z_N^I(0)$ if $N \to +\infty$, as conjectured after equation (20);

(ii) $S_1(N,T) + S_2(N,T) \to 0$ (*cf* equation (A.5)) i.e. $Z_N(0) \sim Z_N^I(0) \sim (4\pi)^{8N^2} u_{\max}^{4N^2}$, with $L = l_{\max} = l$ in equation (27).

As the reasoning is valid for any $[T_{l_i,<}, T_{l_i,>}]$ we finally have in the general case $J_1 \neq J_2$

$$Z_N(0) = (4\pi)^{8N^2} \sum_{l=0}^{+\infty}{}^t \left[f_{l,l} \lambda_l(-\beta J_1) \lambda_l(-\beta J_2)\right]^{4N^2}, \text{ as } N \to +\infty. \quad (29)$$

In the previous equation the special notation $\sum_{l=0}^{+\infty}{}^t$ recalls that the summation can be truncated due to the fact that each eigenvalue $\lambda_l(-\beta J_k)$ is exclusively dominant within the range $[T_{l,<}, T_{l,>}]$. But, if considering the whole temperature range all the $l$-eigenvalues must be kept.

We must also note that, if dealing with a distribution of constant exchange energies $J_1$ and $J_2$ characterizing the horizontal and vertical lattice bonds, respectively, the infinite lattice can be described by the translation of these bonds along the horizontal and vertical axes of the lattice in the crystallographic space. As a result, if using a similar reasoning as the one used for expressing $Z_N(0)$, we can define a zero-field partition function per lattice site symbolically written $z_N(0) = Z_N(0)^{1/4N^2}$ with

$$z_N(0) = (4\pi)^2 \sum_{l=0}^{+\infty}{}^t f_{l,l} \lambda_l(\beta|J_1|) \lambda_l(\beta|J_2|), \text{ as } N \to +\infty. \quad (30)$$

## 3. Spin correlations and correlation length

*3.1 Definitions*

We first define the spin-spin correlation

$$<S_{i,j}.S_{i+k,j+k'}> = \frac{1}{Z_N(0)}\int dS_{-N,-N}...\int dS_{i,j}S_{i,j}...\int dS_{i+k,j+k'}S_{i+k,j+k'}\times...\times$$

$$\times\int dS_{N,N}\exp\left(-\beta\sum_{i=-(N-1)}^{N}\sum_{j=-(N-1)}^{N}H_{i,j}^{ex}\right). \quad (31)$$

In the thermodynamic limit ($N \to +\infty$) $Z_N(0) \sim Z_N^I(0)$ (cf equation (29)). As a result the characteristic polynomial giving the numerator of $<S_{i,j}.S_{i+k,j+k'}>$ is derived from $Z_N^I(0)$ in which $l_{i,j}=l'_{i,j}=l$, $m_{i,j}=m'_{i,j}=0$.

The zero-field spin correlation $<S_u>$, with $u = (i,j)$ or $(i+k,j+k')$ can be obtained from equation (31) by replacing $S_{i+k,j+k'}$ or $S_{i,j}$ by unity. As we deal with isotropic (Heisenberg) couplings, we have the following properties for the site $u = (i,j)$ or $(i+k,j+k')$:

$$<S_{i,j}^v.S_{i+k,j+k'}^v> = \frac{1}{3}<S_{i,j}.S_{i+k,j+k'}>, \quad v = x, y \text{ or } z, <S_u^v> = \frac{1}{\sqrt{3}}<S_u>. \quad (32)$$

The correlation function $\Gamma_{k,k'}$ is:

$$\Gamma_{k,k'} = <S_{i,j}.S_{i+k,j+k'}> - <S_{i,j}><S_{i+k,j+k'}> \quad (33)$$

if $(i,j)$ is the site of reference. In this article we choose $(0,0)$. In addition, due to the isotropic nature of couplings, we have $\Gamma_{k,k'}^v = \Gamma_{k,k'}/3$, $v = x, y$ or $z$.

The general definition of the correlation length is:

$$\xi = \left(\frac{\sum_k\sum_{k'}(k^2+k'^2)|\Gamma_{k,k'}|}{\sum_k\sum_{k'}|\Gamma_{k,k'}|}\right)^{1/2}. \quad (34)$$

Along a horizontal lattice line $k = 0$ ($x$-crystallographic axis of the lattice) $\xi = \xi_x$ (respectively, $k' = 0$ and $\xi = \xi_y$ for a vertical lattice row, $y$-crystallographic axis of the lattice).

Using the general definition of the spin-spin correlation given by equation (31) and expanding the exponential part of the integrand on the infinite basis of spherical harmonics (cf equation (7)), we can write:

$$\begin{pmatrix}<S_{0,0}^z>,<S_{k,k'}^z>\\<S_{0,0}^z.S_{k,k'}^z>\end{pmatrix} = \frac{(4\pi)^{8N^2}}{Z_N(0)}\prod_{i=-(N-1)}^{N}\prod_{j=-(N-1)}^{N}\sum_{l_{i,j},l'_{i,j}}\sum_{m_{i,j},m'_{i,j}}F'_{i,j}\lambda_{l_{i,j}}(-\beta J_1)\lambda_{l'_{i,j}}(-\beta J_2) \quad (35)$$

where $F'_{i,j}$ is the following current integral

$$F'_{i,j} = \int dS_{i,j}X_{i,j}Y_{l'_{i+1,j},m'_{i+1,j}}(S_{i,j})Y_{l_{i,j-1},m_{i,j-1}}(S_{i,j})Y_{l_{i,j},m_{i,j}}^*(S_{i,j})Y_{l'_{i,j},m'_{i,j}}^*(S_{i,j}) \quad (36)$$

for any site $(i,j)$ (and a similar expression for site $(i+k,j+k')$). When $X_{i,j}=1$, $F'_{i,j}=F_{i,j}$ (cf equation (10)). Thus, if calculating $<S_{0,0}^z>$ (or $<S_{k,k'}^z>$) we have a single integral $F'_{0,0}$ (or $F'_{k,k'}$) containing $S_{0,0}^z = X_{0,0} = \cos\theta_{0,0}$ (or $S_{k,k'}^z = X_{k,k'} = \cos\theta_{k,k}$) whereas for $<S_{0,0}^z.S_{k,k'}^z>$ we have two integrals $F'_{0,0}$ and $F'_{k,k'}$ in the product of integrals appearing in equation (35).

### 3.2 Calculation of the spin correlation $<S_u^z>$; consequences

In this subsection we wish to calculate the numerator of the spin correlation $<S_u^z>$. It is given by equations (35) and (36) in which we have $X_{k_1,k_2}=1$ except at the current site $u = (i,j)$ or $u = (i+k,j+k')$ where we use the following recursion relation:

$$\cos\theta_{i,j}Y_{l_{i,j},m_{i,j}}(S_{i,j}) = C_{l_{i,j}+1}Y_{l_{i,j}+1,m_{i,j}}(S_{i,j}) + C_{l_{i,j}-1}Y_{l_{i,j}-1,m_{i,j}}(S_{i,j}) \quad (37)$$

with here $m_{i,j} = 0$ and $m'_{i,j} = 0$ for any site $(i,j)$, in the thermodynamic limit ($N \to +\infty$). Then $C_{l_{i,j}+1}$ and $C_{l_{i,j}-1}$ reduce to

$$C_{l_{i,j}+1} = \frac{l_{i,j}+1}{\sqrt{(2l_{i,j}+1)(2l_{i,j}+3)}} \;,\; C_{l_{i,j}-1} = \frac{l_{i,j}}{\sqrt{(2l_{i,j}+1)(2l_{i,j}-1)}} \;. \tag{38}$$

In the particular case $l_{i,j} = 0$ which occurs at the beginning of each $l$-series expansion, we have $C_{+1} = 1/\sqrt{3}$ and $C_{-1} = 0$. For the calculation of the spin correlation this transform can be equivalently applied to each of the four spherical harmonics appearing in $F'_{i,j}$ given by equation (36). For instance, if wishing to calculate $< S^z_{i,j} >$, we directly apply equation (37) to $Y^*_{l_{i,j},0}(\boldsymbol{S}_{i,j}) = Y_{l_{i,j},0}(\boldsymbol{S}_{i,j})$. As a result $F'_{i,j}$ can be written as:

$$F'_{i,j} = C_{l_{i,j}+1} f_{l_{i,j},l_{i,j}+1} + C_{l_{i,j}-1} f_{l_{i,j},l_{i,j}-1} \tag{39}$$

with:

$$f_{l_{i,j},l_{i,j}+\varepsilon} = \int d\boldsymbol{S}_{i,j} Y_{l_{i+1,j},0}(\boldsymbol{S}_{i,j}) Y_{l_{i,j-1},0}(\boldsymbol{S}_{i,j}) Y_{l_{i,j}+\varepsilon,0}(\boldsymbol{S}_{i,j}) Y_{l'_{i,j},0}(\boldsymbol{S}_{i,j}), \varepsilon = \pm 1, \tag{40}$$

with $l_{i+1,j} = l_{i,j-1} = l_{i,j} = l'_{i,j} = l$. We immediately retrieve the calculation of integrals appearing in that of the zero-field partition function. As a result, if using the expansion of any product of two spherical harmonics given by equation (11) and their orthogonality condition in $f_{l_{i,j},l_{i,j}+\varepsilon}$, we can readily write $f_{l_{i,j},l_{i,j}+\varepsilon} = f_{l,l+\varepsilon}$ i.e.,

$$f_{l,l+\varepsilon} = \frac{(2l+1)^{3/2}(2(l+\varepsilon)+1)^{1/2}}{4\pi} \sum_{L=1}^{\min(2l,2l+\varepsilon)} \frac{1}{2L+1} \left[ C^{L\;0}_{l\;0\;l\;0} C^{L\;0}_{l+\varepsilon\;0\;l\;0} \right]^2, \varepsilon = \pm 1. \tag{41}$$

The non-vanishing condition of the current integral $f_{l,l+\varepsilon}$ which is due to that of the involved C.G. coefficients allows one to write down a *universal temperature-independent selection rule* concerning integers $l$ (cf equation (19)). We now have $C^{L\;0}_{l+\varepsilon\;0\;l\;0} \neq 0$ (with $\varepsilon = \pm 1$) and $C^{L\;0}_{l\;0\;l\;0} \neq 0$ if respectively:

$$2l + \varepsilon + L = 2g \;,\; 2l + L = 2g'. \tag{42}$$

Reporting the $L$-value derived from the first equation i.e., $L = 2g - (2l + \varepsilon)$ in the second one, we have $2g - \varepsilon = 2g'$. The unique solution is $\varepsilon = 0$ which is impossible in the present case because $\varepsilon = \pm 1$, exclusively. As a result $C^{L\;0}_{l+\varepsilon\;0\;l\;0}$ and $C^{L\;0}_{l\;0\;l\;0}$ do not vanish simultaneously but their product is always null. We immediately derive $f_{l,l+\varepsilon} = 0$ ($\varepsilon = \pm 1$) and $F'_{i,j} = 0$ so that $< S^z_{k,k'} > = \pm 1$ for $T = 0$ K, $< S^z_{k,k'} > = 0$ for $T > 0$ K and consequently

$$< \boldsymbol{S}_{k,k'} > = 0, \; \Gamma_{k,k'} = < \boldsymbol{S}_{0,0} . \boldsymbol{S}_{k,k'} >, \; T > 0 \text{ K}. \tag{43}$$

*This result rigorously proves that the critical temperature is absolute zero i.e., $T_c = 0$ K.*

### 3.3 Spin-spin correlation between any couple of lattice sites

In the thermodynamic limit ($N \to +\infty$) on which we exclusively focus, if considering the thermodynamic functions of interest, they are all obtained by deriving $Z_N(0) \sim Z^I_N(0)$ with respect to the temperature $T$ (specific heat) or $Z_N(B)$ with respect to the magnitude of the applied external induction $B$ in the vanishing $B$-limit (spin-spin correlations, correlation length and susceptibility). As a result the numerator of *all these functions show the same l-polynomial structure as $Z_N(0)$ which appears at their denominator.*

For each *l*-term, the angular factor is such as $m_{k,k'} = m'_{k,k'} = 0$, for any lattice site. The radial factor is similar to that of $Z_N(0)$ i.e., $[f_{l,l}\lambda_l(-\beta J)]^{4N^2}$ multiplied by another factor $\Phi_{\alpha(l)}(-\beta J)^K$ coming from the adequate derivation of $Z_N(0)$ with respect to $T$ or $Z_N(B)$ with respect to $B$ as $B \to 0$. $\alpha(l)$ and $K$ are characteristics of the thermodynamic function.

As a result the numerator of each of these thermodynamic functions is directly expressed as a new characteristic *l*-polynomial characterized by a new set of *l*-eigenvalues $[f_{l,l}\lambda_l(-\beta J)]^{4N^2}\Phi_{\alpha(l)}(-\beta J)^K$. It means that, as for $Z_N(0)$, *there also exist thermal crossovers between these new l-eigenvalues. Their respective thermal domains of predominance are not necessary the same ones as those of eigenvalues* $\lambda_l(-\beta J)$ *appearing in* $Z_N(0)$. $\Phi_{\alpha(l)}(\beta|J|)^K$, $\alpha(l)$ and $K$ will be identified below in the case of spin-spin correlations (*cf* equation (48c)).

The *z-z* spin-spin correlation $< S^z_{0,0} \cdot S^z_{k,k'} >$ is given by equations (31) and (35). We restrict the following study to $k > 0$ and $k' > 0$, without loss of generality. Due to the presence of $\cos\theta_{0,0}$ and $\cos\theta_{k,k'}$ appearing in the integrals $F'_{0,0}$ and $F'_{k,k'}$ which characterize the spin orientations at sites $(0,0)$ and $(k,k')$ we have to reconsider a new integration process. This process is similar to that one used for calculating $Z_N(0)$. It can be mainly carried out through two methods:

(i) integrating simultaneously over all the sites from the four lattice lines $i = N$, $i = -(N-1)$, $j = N$ and $j = -(N-1)$ in the direction of the lattice heart;

(ii) integrating from horizontal line $i = -(N-1)$ to $i = N$ between vertical lines $j = -(N-1)$ and $j = N$ (lines $i$ or $j = N$ and $i$ or $j = -N$ being confused on the torus, respectively) or *vice versa*.

It is useful to combine both methods. In a first step we choose method (i). In the dominant *l*-term the integrals $F_{i,j}$ involving sites located far from correlated sites $(0,0)$ and $(k,k')$ are characterized by a collection of integers $l'_{i+1,j} = l_{i,j-1} = l_{i,j} = l'_{i,j} = l$ for reasons explained in Subsec. 2.5. This part of the lattice constitutes the *wing domain*. When reaching the horizontal lattice lines $i = 0$ and $k$ and the vertical ones $j = 0$ and $k'$ whose respective intersections two by two are sites $(0,0)$, $(0,k')$, $(k,k')$ and $(k,0)$ a special care must be brought. The inner domain defined by these two couples of lines is *the correlation domain*.

A consequence of method (i) is that all the bonds located outside the correlation domain (or *out-bonds*) are characterized by the integer $l$, notably all the bonds linked to the frontier of the correlation domain. All the previous results are summarized in the following theorem:

**Theorem 1 (confinement theorem)**

*In the thermodynamic limit, for calculating the numerator of the spin-spin correlation* $< S^z_{0,0} \cdot S^z_{k,k'} >$, *it is necessary to take into account two domains: a correlation domain which is a rectangle of vertices* $(0,0)$, $(0,k')$, $(k,k')$ *and* $(k,0)$ *within which all the correlation paths are confined, and a remaining domain called wing domain. In both domains, for an infinite lattice, we have* $m = 0$. *All the bonds of the wing domain are characterized by the same integer l, including the bonds linked to the correlation domain.*

In a second step we use the integration method (ii) i.e., line by line. *The decomposition law given by equation* (37) *only intervenes at the correlated sites* $(0,0)$ *and* $(k,k')$ for which we have $l_{0,-1} = l'_{0,0} = l$ and $l'_{k+1,k'} = l_{k,k'} = l$, respectively, due to the integration in the wing domain. The corresponding integrals characterizing these sites are $F'_{0,0}$ and $F'_{k,k'}$ given by equations (36), (39) and (40). The other integrals describing the spin states of the current sites $(i,j)$ involved in the integration process inside the correlation domain are $F_{i,j}$-like given by equation (10) except for sites belonging to the correlation path.

First let us consider for instance site $(0,0)$. The integrand of integral $F'_{0,0}$ given by equation (36) is $\cos\theta_{0,0}[Y_{l,0}(S_{0,0})]^2 Y_{l_{0,0},0}(S_{0,0})Y_{l'_{1,0},0}(S_{0,0})$ as $l_{0,0}$ and $l'_{1,0}$ characterize bonds of the correlation domain not yet examined through the integration process. The decomposition law can be applied to the spherical harmonics $Y_{l,0}(S_{0,0})$, $Y_{l_{0,0},0}(S_{0,0})$ or $Y_{l'_{1,0},0}(S_{0,0})$. The integrand of $F'_{0,0}$ becomes $Y_{l,0}(S_{0,0})Y_{l\pm1,0}(S_{0,0})Y_{l_{0,0},0}(S_{0,0})Y_{l'_{1,0},0}(S_{0,0})$ (integrand 1), $[Y_{l,0}(S_{0,0})]^2 Y_{l_{0,0}\pm1,0}(S_{0,0})Y_{l'_{1,0},0}(S_{0,0})$ (integrand 2) and $[Y_{l,0}(S_{0,0})]^2 \times Y_{l_{0,0},0}(S_{0,0})Y_{l'_{1,0}\pm1,0}(S_{0,0})$ (integrand 3), respectively. We immediately retrieve the calculation of integrals appearing in that of the zero-field partition function. We express the products of pairs of spherical harmonics as C.G. series (*cf* equation (11)). For instance, if examining integrand 1, we can use the following combinations

$$Y_{l,0}(S_{0,0})Y_{l'_{1,0},0}(S_{0,0}) = \sum_{L=|l-l'_{1,0}|}^{l+l'_{1,0}} \left[\frac{(2l+1)(2l'_{1,0}+1)}{4\pi(2L+1)}\right]^{1/2} \left[C_{l\ 0\ l'_{1,0}\ 0}^{L\ 0}\right]^2 Y_{L,0}(S_{0,0}), \qquad (44)$$

$$Y_{l_{0,0},0}(S_{0,0})Y_{l\pm1,0}(S_{0,0}) = \sum_{L'=|l_{0,0}-(l\pm1)|}^{l_{0,0}+l\pm1} \left[\frac{(2l_{0,0}+1)(2(l\pm1)+1)}{4\pi(2L'+1)}\right]^{1/2} \left[C_{l_{0,0}\ 0\ l\pm1\ 0}^{L'\ 0}\right]^2 Y_{L',0}(S_{0,0}).$$

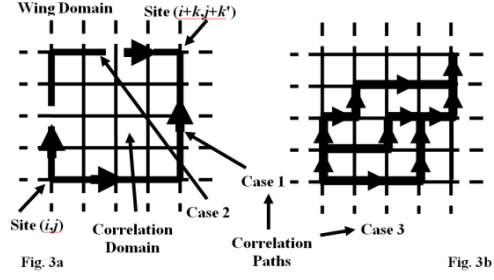

**Figure 3**. a) Correlation paths (thick lines) along the frontier between the correlation and the wing domains (cases 1 and 2); for a given path all the bonds are characterized by $l+1$ or $l-1$; the other bonds not involved in the correlation path (thin lines) are characterized by $l$; b) examples of correlation paths inside the correlation domain (case 3).

Introducing these C.G. series in integral $F'_{0,0}$ given by equation (36) and using the orthogonality condition of spherical harmonics leads to $L = L'$ i.e., notably $L_< = L'_<$ and $L_> = L'_>$. Recalling that the characteristic polynomial associated with the numerator of $<S^z_{0,0} \cdot S^z_{k,k'}>$ is derived from $Z_N(0) \sim Z_N^I(0)$ where $l'_{i+1,j} = l_{i,j-1} = l_{i,j} = l'_{i,j} = l$ for any site $(i,j)$ the unique solutions are $l_{0,0} = l \pm 1$, $l'_{1,0} = l$ (case 1) and $l_{0,0} = l$, $l'_{1,0} = l \pm 1$ (case 2). All the other combinations between pairs of spherical harmonics lead to the same couple of solutions $l_{0,0}$ and $l'_{1,0}$, for integrand 1 but also for integrands 2 and 3. From a mathematical point of view it also means that, for case 1 or 2, *there are only two channels of integration leading to: a path beginning with a bond such as $l_{0,0} = l + 1$ and another one with $l_{0,0} = l - 1$ (case 1) or a path with $l'_{1,0} = l + 1$ and another one with $l'_{1,0} = l - 1$ (case 2)*.

(i) Case 1 (see Fig. 3a)

We apply the decomposition law to the spherical harmonics $Y_{l,0}(S_{0,0})$ (integrand 1). We choose $l'_{1,0} = l$. Integral $F'_{0,0}$ given by equation (39) can be written as

$$F'_{0,0} = C_{l+1} f_{l_{0,0},l+1} + C_{l-1} f_{l_{0,0},l-1}, \qquad (45a)$$

where $C_{l\pm 1}$ is defined by equation (38) and with:

$$f_{l_{0,0},l+\varepsilon} = \int dS_{0,0} [Y_{l,0}(S_{0,0})]^2 Y_{l_{0,0},0}(S_{0,0}) Y_{l+\varepsilon,0}(S_{0,0}), \quad \varepsilon = \pm 1. \qquad (45b)$$

As just seen the non-vanishing condition of integral $f_{l_{0,0},l+\varepsilon}$ imposes $l_{0,0} = l + \varepsilon$, $\varepsilon = \pm 1$. Thus all the bonds linked to $(0,0)$ are characterized by the integer $l$ whereas the unique bond of the correlation domain (or *in-bond*) is characterized by $l_{0,0} = l \pm 1$. We have

$$F'^{+}_{0,0} = C_{l+1}f_{l+1,l+1} + C_{l-1}f_{l+1,l-1}, \quad l_{0,0} = l+1,$$
$$F'^{-}_{0,0} = C_{l+1}f_{l-1,l+1} + C_{l-1}f_{l-1,l-1}, \quad l_{0,0} = l-1, \quad (46)$$

with

$$f_{l+\varepsilon',l+\varepsilon} = \frac{(2l+1)[(2(l+\varepsilon)+1)(2(l+\varepsilon')+1)]^{1/2}}{4\pi}\sum_{L=0}^{L_>}\frac{1}{2L+1}\left[C^{L\ 0}_{l\ 0\ l\ 0}\ C^{L\ 0}_{l+\varepsilon'\ 0\ l+\varepsilon\ 0}\right]^2,$$

$$\varepsilon = \pm 1, \varepsilon' = \pm 1, \quad (47)$$

and $L_> = \min(2l, 2l + \varepsilon + \varepsilon')$. In the previous equation $C^{L\ 0}_{l\ 0\ l\ 0}$ does not vanish if $2l + L = 2g$ and $C^{L\ 0}_{l+\varepsilon'\ 0\ l+\varepsilon\ 0}$ if $2l + \varepsilon + \varepsilon' + L = 2g'$. Thus the product of C.G.'s does not vanish if $\varepsilon + \varepsilon' = 2(g' - g)$ i.e., $\varepsilon + \varepsilon' = \pm 2$ ($\varepsilon = \varepsilon' = \pm 1$), $g' - g = \pm 1$ and $\varepsilon + \varepsilon' = 0$ ($\varepsilon = -\varepsilon' = \pm 1$), $g' = g$.

As a result the corresponding contribution of site (0,0) to the numerator of $<S^z_{0,0}.S^z_{k,k'}>$ is $F'^{\pm}_{0,0}\lambda_{l\pm 1}(-\beta J_1)$ where $\lambda_{l\pm 1}(-\beta J_1)$ and $F'^{\pm}_{0,0}$ are respectively given by equations (8) and (45)-(47). In other words, in case 1, the beginning of the correlation path is constituted by the bond between sites (0,0) and (0,1) characterized by $l_{0,0} = l \pm 1$.

Now we consider the other sites of line $i = 0$ i.e., sites (0,1) to (0,k'). At site (0,1) there is *no decomposition law*. We have $l_{0,0} = l \pm 1$ due to integration at site (0,0) and $l'_{0,1} = l$ due to integration in the wing domain. As a result, if examining integral $F_{0,1}$ given by equation (10), the integrand is nothing but $Y_{l,0}(S_{0,1})Y_{l'_{1,1},0}(S_{0,1})Y_{l_{0,1},0}(S_{0,1})Y_{l\pm 1,0}(S_{0,1})$.

As seen after equation (44) all the decompositions of products of spherical harmonics pairs as C.G. series only lead to two possible choices. If $l'_{1,1} = l$ the non-vanishing condition of $F_{0,1}$ (cf equation (47)) imposes $l_{0,1} = l_{0,0} = l \pm 1$ and the correlation path continues along the horizontal line $i = 0$ (case 1). If $l_{0,1} = l$ the non-vanishing condition of $F_{0,1}$ now imposes $l'_{1,1} = l \pm 1$ and the correlation path continues along the vertical line $j = 1$: we then deal with a new correlation path called case 3 and detailed below (see Fig. 3b).

In summary, in case 1, we have two types of correlation path between sites (0,0) and (0,1): the bond is characterized by $l+1$ ($F_{0,1} = f_{l+1,l+1}$) or $l-1$ ($F_{0,1} = f_{l-1,l-1}$). This situation is similar for all the sites of line $i = 0$ i.e., between sites (0,1) and (0,k'−1). As a result the corresponding contribution to the numerator of $<S^z_{0,0}.S^z_{k,k'}>$ is $\left(F'^{\pm}_{0,0}/f_{l\pm 1,l\pm 1}\right)\times$ $\times\left(f_{l\pm 1,l\pm 1}\lambda_{l\pm 1}(-\beta J_1)\right)^{k'}$. In addition, due to the integration process which has swept all the in-bonds of the correlation domain not involved in the correlation path we have $l_{K,K'} = l'_{K,K'} = l$ for all the horizontal and vertical bonds except for those of vertical line $j = k'$.

Arriving at site (0,k') we have to determine $l'_{1,k'}$ because $l_{0,k'} = l'_{0,k'} = l$ due to the wing contribution and $l_{0,k'-1} = l \pm 1$ due to the non-vanishing condition of integral $F_{0,k'-1} = f_{l\pm 1,l\pm 1}$. That of integral $F_{0,k'}$ gives $l'_{1,k'} = l \pm 1$ and $F_{0,k'} = F_{0,k'-1} = f_{l\pm 1,l\pm 1}$. Then the work of integration is similar for the remaining sites of the vertical line $j = k'$ between sites (1,k') and (k−1,k'), with for this later site $l'_{k,k'} = l \pm 1$. The corresponding contribution to the numerator of $<S^z_{0,0}.S^z_{k,k'}>$ is $\left(f_{l\pm 1,l\pm 1}\lambda_{l\pm 1}(-\beta J_2)\right)^k / f_{l\pm 1,l\pm 1}$.

Arriving at site (k,k') the integers $l'_{k+1,k'}$ and $l_{k,k'}$ have been already determined in the wing domain ($l'_{k+1,k'} = l_{k,k'} = l$) or along the correlation path ($l'_{k,k'} = l \pm 1$). Concerning integral $F'^{\pm}_{k,k'}$ an independent study similar to that achieved at site (0,0) can be done. We have $F'^{\pm}_{k,k'} = F'^{\pm}_{0,0}$ where the integral $F'^{\pm}_{0,0}$ is given by equations (45a)-(47). Here the unique solutions are $l'_{k,k'} = l \pm 1$, $l_{k,k'-1} = l$ (case 1) and $l'_{k,k'} = l$, $l_{k,k'-1} = l \pm 1$ (case 2). The final contribution of all the sites to the correlation path between sites (0,0) and (k,k') is $\left(F'^{\pm}_{0,0}\right)^2\left(f_{l\pm 1,l\pm 1}\lambda_{l\pm 1}(-\beta J_1)\right)^{k'}\left(f_{l\pm 1,l\pm 1}\lambda_{l\pm 1}(-\beta J_2)\right)^k / f_{l\pm 1,l\pm 1}$ (case 1).

In summary all the horizontal bonds of the correlation path are characterized by $l_{0,K'} = l \pm 1$ ($0 \leq K' \leq k'-1$) between sites (0,0) and (0,k') on the one hand and all the vertical bonds by $l'_{K,k'} = l \pm 1$ ($1 \leq K \leq k$) between sites $(0,k')$ and $(k,k')$ on the other one. All the bonds of the correlation domain not involved in the correlation path are characterized by the integer $l$.

(ii) Case 2 (see Fig. 3a)

Now we impose $l_{0,0} = l$. The work is strictly similar but the correlation path concerns the vertical bonds between sites $(0,0)$ and $(k,0)$ for which $l'_{K,0} = l \pm 1$ ($1 \leq K \leq k$) and the horizontal ones between sites $(k,0)$ and $(k,k')$ for which $l_{k,K'} = l \pm 1$ ($0 \leq K' \leq k'-1$).

(iii) Case 3 (see Fig. 3b)

This case is a mix of cases 1 and 2. At each current site $(i,j)$ belonging to the correlation domain we can have $l_{i,j} = l \pm 1$, $l'_{i+1,j} = l$ or $l'_{i+1,j} = l \pm 1$, $l_{i,j} = l$.

*Due to the fact that it is impossible to go backwards when the integration process has been carried out any loop of the correlation path is forbidden. In addition the expression of the numerator of the spin-spin correlation is independent of the bond orientation chosen for the integration process i.e., between sites $(0,0)$ and $(k,k')$ or vice versa.*

We conclude that

(i) all the correlation paths contain the same number of horizontal and vertical bonds; as a result all these paths show the same expression i.e., all the spin-spin correlations show a unique expression, as expected for this kind of lattice;

(ii) *these correlation paths correspond to the shortest possible length through the bonds involved between sites $(0,0)$ and $(k,k')$;* their total number is simply $n = \binom{k+k'}{k'}$; thus there are $n_{l+1} = n$ paths whose bonds are characterized by the integer $l+1$ and $n_{l-1} = n$ paths showing bonds characterized by $l-1$; they have the same weight $w_{l+1} = w_{l-1} = n_{l\pm1}/2n = 1/2$.

**Theorem 2**

*As loops are forbidden for all the correlation paths these paths have the same length inside the correlation domain. This length is the shortest possible one through the lattice bonds between any couple of correlated sites. Each path respectively involves the same number of horizontal and vertical bonds as the horizontal and vertical sides of the correlation rectangle, for a 2d-infinite square lattice.*

As a result the spin-spin correlation $<S^z_{0,0}.S^z_{k,k'}>$ can be written:

$$<S^z_{0,0}.S^z_{k,k'}> = \frac{(4\pi)^{8N^2}}{2Z_N(0)} \sum_{l=0}^{+\infty}{}^t z_l^{4N^2} \left[ K^2_{l+1}\left(P_{1,l+1}\right)^{k'}\left(P_{2,l+1}\right)^{k} + K^2_{l-1}\left(P_{1,l-1}\right)^{k'}\left(P_{2,l-1}\right)^{k} \right],$$

$$k > 0, k' > 0, \text{ as } N \to +\infty \quad (48a)$$

where

$$z_l = f_{l,l}\lambda_l(\beta|J_1|)\lambda_l(\beta|J_2|), \quad K_{l\pm1} = \sqrt{\frac{f_{l,l}}{f_{l\pm1,l\pm1}}}\left( C_{l+1}\frac{f_{l\pm1,l+1}}{f_{l,l}} + C_{l-1}\frac{f_{l\pm1,l-1}}{f_{l,l}} \right),$$

$$Z_N(0) = \sum_{l=0}^{+\infty} z_l^{4N^2}, \quad P_{i,l\pm1} = \frac{f_{l\pm1,l\pm1}}{f_{l,l}} \frac{\lambda_{l\pm1}(-\beta J_i)}{\lambda_l(-\beta J_i)}, \quad i = 1, 2. \quad (48b)$$

The integrals $f_{l,l}$ and $f_{l\pm1,l\pm1}$ are given by equations (20) and (47). In the previous equation the special notation $\sum_{l=0}^{+\infty}{}^t$ has been defined after equation (29) but here it concerns the new $l$-eigenvalues $P_{i,l\pm1}$. Finally, owing to equation (32) one can express $<S_{0,0}.S_{k,k'}>$.

As explained at the beginning of this subsection the current term of the spin-spin correlation has the announced form $\left[f_{l,l}\lambda_l(-\beta J)\right]^{4N^2} \Phi_{\alpha(l)}(-\beta J)^K$. When $J_1 = J_2$

$$\Phi_{l\pm1}(-\beta J)^K = K^2_{l\pm1}\Theta_{l\pm1}(-\beta J)^{k+k'}, \quad \Theta_{l\pm1}(-\beta J) = P_{l\pm1}, \quad \alpha(l) = l \pm 1, \quad K = k + k' \quad (48c)$$

and a similar but functional equation when $J_1 \neq J_2$.

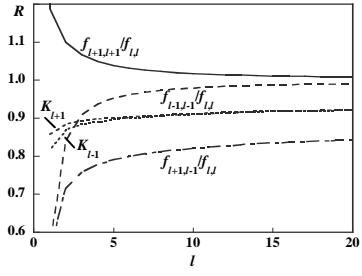 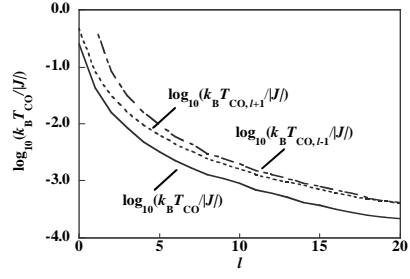

**Figure 4**. Plots of the ratios $R = f_{l+1,l+1}/f_{l,l}$, $f_{l-1,l-1}/f_{l,l}$, and $f_{l+1,l-1}/f_{l,l}$ where integrals $f_{l+\varepsilon,l+\varepsilon'}$ ($\varepsilon = \pm 1$, $\varepsilon' = \pm 1$) are defined by equations (10), (20), (45b) and (47) as well quantities $K_{l+1}$ and $K_{l-1}$ given by equation (48b).

**Figure 5**. Plots of $\log_{10}(k_B T_{CO}/|J|)$, $\log_{10}(k_B T_{CO,l+1}/|J|)$ and $\log_{10}(k_B T_{CO,l-1}/|J|)$; the crossover temperatures $T_{CO}$, $T_{CO,l+1}$ and $T_{CO,l-1}$ are defined by the transcendental equations (26) and (50), respectively.

### 3.4 Properties of spin-spin correlation

Due to the classical nature of spin momenta (cf equation (6)), we have seen that the full lattice operator $\exp(-\beta H^{ex})$ can be written as the product $\exp(-\beta H^{ex,H}) \times \exp(-\beta H^{ex,V})$ where $\exp(-\beta H^{ex,H})$ and $\exp(-\beta H^{ex,V})$ are the respective operators of the set of horizontal and vertical lattice lines. As a result each term of the $l$-summation giving $Z_N(0)$ i.e., each $l$-eigenvalue, appears as the product of the corresponding eigenvalues $\lambda_l(-\beta J_H)\lambda_l(-\beta J_V)$. This property also exists for the spin-spin correlation. Indeed, if examining the closed-form expression given by equation (48a), it can be written

$$<S_{0,0}.S_{k,k'}> = \frac{1}{2Z_N(0)} \sum_{l=0}^{+\infty} {}^t \sum_{\varepsilon=\pm 1} z_l^{4N^2} <S_{0,0}.S_{k,0}>_{l+\varepsilon} <S_{0,0}.S_{0,k'}>_{l+\varepsilon}, \quad (49)$$

with $<S_{0,0}.S_{u,v}>_{l+\varepsilon} = fK_{l+\varepsilon}(P_{i,l+\varepsilon})^\alpha$, $f = \sqrt{3}$, $i = 1$, $u = 0$, $v = \alpha = k'$, $i = 2$, $u = \alpha = k$, $v = 0$. The factor $z_l^{4N^2}/Z_N(0) = z_l^{4N^2}/\sum_l z_l^{4N^2}$ appears as the weight of the $l$-state.

When $d = 1$ ($D = 2$) there is no path characterized by $l-1$ and no $l$-summation (in the thermodynamic limit the $l$-series is restricted to $l = 0$) so that equation (49) reduces to $<S_0.S_k> = <S_0.S_u><S_{u+1}.S_k>$. In the case $d = 2$ ($D = 3$) this property only concerns the $l$-current term $<S_{0,0}.S_{k,0}>_{l+\varepsilon} = <S_{0,0}.S_{u,0}>_{l+\varepsilon} <S_{u+1,0}.S_{k,0}>_{l+\varepsilon}$ and $<S_{0,0}.S_{0,k'}>_{l+\varepsilon} = <S_{0,0}.S_{0,v}>_{l+\varepsilon} <S_{0,v+1}.S_{0,k'}>_{l+\varepsilon}$ i.e., for each $l$-state of the whole lattice.

As a result, if considering the susceptibility $\chi = \beta G \sum_k \sum_{k'} <S_{0,0}.S_{k,k'}>$ of a lattice composed of spin momenta showing the same Landé factor $G$, one can predict that it can be put under the form $\chi = (2Z_N(0))^{-1} \sum_{l=0}^{+\infty} \sum_{\varepsilon=\pm 1} z_l^{4N^2} \chi_{l+\varepsilon}$, with $\chi_{l+\varepsilon} = \chi_{l+\varepsilon}^H \chi_{l+\varepsilon}^V$ where $\chi_{l+\varepsilon}^H$ (respectively, $\chi_{l+\varepsilon}^V$) is the susceptibility of the full horizontal (respectively, vertical) lattice lines. The study of $\chi$ is out of the present article framework.

Now we have to examine the ratios $P_{i,l+1}$ and $P_{i,l-1}$ ($i = 1,2$) defined by equation (48b). For physical reasons we must have $|P_{i,l\pm 1}| \leq 1$. In Fig. 4 we have reported the various ratios of integrals $f_{l+1,l+1}/f_{l,l}$, $f_{l-1,l-1}/f_{l,l}$ and $f_{l+1,l-1}/f_{l,l}$. We have $f_{l+1,l+1}/f_{l,l} > 1$, but $f_{l-1,l-1}/f_{l,l} < 1$, $f_{l+1,l-1}/f_{l,l} < 1$ as well as $K_{l+1} < 1$ and $K_{l-1} < 1$.

As a result, for a given relative temperature $\beta|J_i| = |J_i|/k_B T$ ($i = 1,2$), we always have $I_{l+3/2}(\beta|J_i|) < I_{l+1/2}(\beta|J_i|)$, with $f_{l+1,l+1}/f_{l,l} > 1$, so that $|P_{i,l+1}| < 1$ or $|P_{i,l+1}| > 1$ which has no

physical meaning. Similarly we have $I_{l-1/2}(\beta|J_i|) > I_{l+1/2}(\beta|J_i|)$, with $f_{l-1,l-1}/f_{l,l} < 1$, and we can deal with $|P_{i,l-1}| < 1$ or $|P_{i,l-1}| > 1$.

We proceed as for the thermal study of the current term of the $l$-polynomial expansion of $Z_N(0)$ where we have defined a crossover temperature $T_{CO}$ through equation (26) so that the eigenvalue $\lambda_{l_i}(\beta|J_i|)$ is dominant within the range $[T_{l_i,<}, T_{l_i,>}]$ and becomes subdominant outside this range. Similarly, for studying the ratio $\lambda_{l\pm1}(\beta|J|)/\lambda_l(\beta|J|)$ appearing in $\Phi_{\alpha(l)}(\beta|J_i|) = |P_{i,l\pm1}|$, we impose $\max(|P_{i,l\pm1}|) = 1$. As a result, from equation (48b) we respectively define two new crossover temperatures $T_{CO,l+1}$ and $T_{CO,l-1}$ by

$$\frac{\lambda_{l+1}(|J|/k_B T_{CO,l+1})}{\lambda_l(|J|/k_B T_{CO,l+1})} = \frac{f_{l,l}}{f_{l+1,l+1}}, \quad \frac{\lambda_{l-1}(|J|/k_B T_{CO,l-1})}{\lambda_l(|J|/k_B T_{CO,l-1})} = \frac{f_{l,l}}{f_{l-1,l-1}} \tag{50}$$

in the simplest case $J = J_1 = J_2$ without loss of generality. It can be extended to the case $J_1 \neq J_2$. As for equation (26) we have numerically solved this transcendental equation.

We find that $|P_{l-1}| \leq 1$ if $T \leq T_{CO,l-1}$ and $|P_{l-1}| > 1$ if $T > T_{CO,l-1}$ on the one hand but $|P_{l+1}| \geq 1$ if $T \leq T_{CO,l+1}$ and $|P_{l+1}| < 1$ if $T > T_{CO,l+1}$ on the other one. Thus, this is the competition between the smooth decreasing $l$-law of the ratio $f_{l+1,l+1}/f_{l,l} > 1$ (see Fig. 4) which tends towards unity when $l \to +\infty$ (i.e., when $T$ tends to $T_c = 0$ K) and the $T$-law of the ratio $\lambda_{l+1}(\beta|J|)/\lambda_l(\beta|J|) < 1$ involved in $|P_{l+1}|$ which is responsible of such a crossover. For $|P_{l-1}|$ the competition is between the increasing $l$-law of the ratio $f_{l-1,l-1}/f_{l,l} < 1$ and the $T$-law of the ratio $\lambda_{l-1}(\beta|J|)/\lambda_l(\beta|J|) > 1$.

Finally, owing to the numerical study reported in Fig. 5, we have $T_{CO} < T_{CO,l+1} < T_{CO,l-1}$. As a result it becomes possible to determine the new domains of thermal predominance $[T_{l\pm1,<}, T_{l\pm1,>}]$ of the eigenvalues $|P_{l\pm1}|$. Their detailed classification is out of the framework of the present article. When $l \to +\infty$ i.e., as $T$ approaches $T_c = 0$ K, all the $l$-eigenvalues become equivalent but a common limit can be selected.

### 3.5 Correlation length

The correlation length can be derived owing to equation (34). We have

$$\xi = \left[\frac{\sum_{l=0}^{+\infty} {}^t z_l^{4N^2}\left[(N_{l+1}^x + N_{l-1}^x) + (N_{l+1}^y + N_{l-1}^y)\right]}{\sum_{l=0}^{+\infty} {}^t z_l^{4N^2}[D_{l+1} + D_{l-1}]}\right]^{1/2} = \sqrt{\xi_x^2 + \xi_y^2}, \text{ as } N \to +\infty \tag{51a}$$

with on condition that $|P_{i,l\pm1}| \leq 1$

$$N_{l\pm1}^x + N_{l\pm1}^y = 2D_{l\pm1}\left[\frac{|P_{1,l\pm1}|}{(1-|P_{1,l\pm1}|)^2} + \frac{|P_{2,l\pm1}|}{(1-|P_{2,l\pm1}|)^2}\right], \quad D_{l\pm1} = K_{l\pm1}^2 \frac{(1+|P_{1,l\pm1}|)(1+|P_{2,l\pm1}|)}{(1-|P_{1,l\pm1}|)(1-|P_{2,l\pm1}|)}. \tag{51b}$$

$z_l$, $P_{i,l+\varepsilon}$ ($i = 1,2$) $K_{l+\varepsilon}$ and the integral $f_{l,l}$ appearing in $z_l$ are respectively given by equations (48b) and (20) in which $l_i = l_j = l$, $m = 0$. $K_{l+\varepsilon} \to 1$ as $l \to +\infty$.

In equation (51b), if considering the spin-spin correlation between first-nearest neighbours derived from equation (48a) and setting $|<S_{0,0}^z \cdot S_{0,1}^z>_{l\pm1}| = K_{l\pm1}|P_{1,l\pm1}|$, $|<S_{0,0}^z \cdot S_{1,0}^z>_{l\pm1}| = K_{l\pm1}|P_{2,l\pm1}|$, the $l$-contribution to the spin-spin correlation is $|<S_{0,0}^z \cdot S_{0,1}^z>|_l = f(K_{l+1}|P_{1,l+1}| + K_{l-1}|P_{1,l-1}|)/2$ ($|<S_{0,0}^z \cdot S_{1,0}^z>|_l$ is derived from $|<S_{0,0}^z \cdot S_{0,1}^z>|_l$ by exchanging 1 against 2); the factor $f$ given after equation (49) can be omitted in equation (51a). As a result we can write

$$N_{l+1}^x + N_{l-1}^x = \frac{N_l^x}{(D_l^x)^3 D_l^y}, \quad D_{l+1}^x + D_{l-1}^x = \frac{N_l^{'x}}{D_l^x D_l^y}, \quad N_{l+1}^y + N_{l-1}^y = \frac{N_l^y}{(D_l^y)^3 D_l^x}, \quad D_{l+1}^y + D_{l-1}^y = \frac{N_l^{'y}}{D_l^x D_l^y},$$
(52a)

with

$$D_l^x = 1 - 2|<S_{0,0}^z.S_{0,1}^z>|_l + |<S_{0,0}^z.S_{0,1}^z>_{l+1}||<S_{0,0}^z.S_{0,1}^z>_{l-1}|,$$
$$D_l^y = 1 - 2|<S_{0,0}^z.S_{1,0}^z>|_l + |<S_{0,0}^z.S_{1,0}^z>_{l+1}||<S_{0,0}^z.S_{1,0}^z>_{l-1}|.$$
(52b)

It is not necessary to express $N_l^x$ and $N_l^y$ because the behaviour of the correlation length $\xi$ near $T_c = 0K$ is essentially ruled by its denominator. Due to its definition (*cf* equations (32)-(34)) $|<S_{0,0}^z.S_{k,k'}^z>|$ can be replaced by $|<S_{0,0}.S_{k,k'}>|$ in equation (51a). $\xi$ *exactly shows the same thermal crossovers as* $|<S_{0,0}.S_{k,k'}>|$. Thus, in the temperature range $[T_{l,<},T_{l,>}]$ where the *l*-eigenvalue of the spin-spin correlation is dominant, we have

$$\xi_x \approx \frac{N_{x,l}}{D_l^x}, \quad \xi_y \approx \frac{N_{y,l}}{D_l^y}, \quad N_{x,l} = \sqrt{\frac{N_l^x}{N_l^{'x}}}, \quad N_{y,l} = \sqrt{\frac{N_l^y}{N_l^{'y}}}, \quad T \in [T_{l,<},T_{l,>}]. \quad (52c)$$

Near $T_c = 0K$, we have previously shown that $l \to +\infty$. Under these conditions $|<S_{0,0}.S_{0,1}>| \approx |<S_{0,0}.S_{0,1}>_{l\pm1}| \approx 1$ and $|<S_{0,0}.S_{1,0}>| \approx |<S_{0,0}.S_{1,0}>_{l\pm1}| \approx 1$ so that

$$\xi_x \approx N_{x,l}/2(1 - |<S_{0,0}.S_{0,1}>|), \quad \xi_y \approx N_{y,l}/2(1 - |<S_{0,0}.S_{1,0}>|), \text{ as } l \to +\infty. \quad (52d)$$

It means that the spin-spin correlation between first-nearest neighbours plays a fundamental role near the critical point. *This is a hidden consequence of equation (49) itself derived from the classical character of spin momenta (cf equation (6)).*

## 4. Low-temperature behaviours

### 4.1 Preliminaries

For sake of simplicity we again reduce the study to the simplest case $J = J_1 = J_2$ without loss of generality. We examine the low-temperature behaviour of the ratios $P_{i,l+\varepsilon}$ involved in the spin-spin correlation (*cf* equations (48a), (48b)), with here $P_{i,l+\varepsilon} = P_{l+\varepsilon}$ ($i = 1,2$).

We first consider the ratio $f_{l+\varepsilon,l+\varepsilon}/f_{l,l}$ ($\varepsilon=\pm1$) where integrals $f_{l,l}$ and $f_{l+\varepsilon,l+\varepsilon}$ are respectively given by equation (47). The *l*-behaviour of each of these ratios has been reported in Fig. 4. $f_{l+1,l+1}/f_{l,l} > 1$ and $f_{l-1,l-1}/f_{l,l} < 1$ but $f_{l\pm1,l\pm1}/f_{l,l} \to 1$ as $l \to +\infty$ i.e., near the critical temperature $T_c = 0$ K. Indeed, if expressing the spherical harmonics involved in the definition of integral $F_{l,l}$ given by equation (10) in which $m = 0$, we have in the infinite *l*-limit [26]

$$Y_{l,0}(\theta,\varphi) \approx \frac{1}{\pi\sqrt{\sin\theta}}\left\{\left(1 - \frac{3}{8l}\right)\cos\left((2l+1)\frac{\theta}{2} - \frac{\pi}{4}\right) - \frac{1}{8l\sin\theta}\cos\left((2l+3)\frac{\theta}{2} - \frac{3\pi}{4}\right)\right\} + O\left(\frac{1}{l^2}\right),$$
as $l \to +\infty$, $\varepsilon' \le \theta \le \pi - \varepsilon'$, $0 < \varepsilon' << 1/l$, $0 \le \varphi \le 2\pi$. (53)

and the exact asymptotic result:

$$\frac{f_{l+\varepsilon,l+\varepsilon}}{f_{l,l}} \to 1 + O\left(\frac{1}{l^2}\right), \quad \varepsilon = \pm1, \text{ as } l \to +\infty. \quad (54)$$

Then, if taking into account equations (8) and (48b), $P_{l\pm1}$ behaves as

$$P_{l\pm1} \approx \frac{I_{l\pm1}(-\beta J)}{I_l(-\beta J)}, \text{ as } T \to 0 \ (l \to +\infty, J_1 = J_2). \quad (55)$$

Intuitively, in the low-temperature limit, we must consider the three cases $\beta|J| >> l$, $\beta|J| \sim l$ and $\beta|J| << l$. The behaviour of the Bessel function $I_l(-\beta|J|)$ in the double limit $l \to +\infty$ and $\beta|J| \to +\infty$ has been established by Olver [27]. In previous papers [24] we have extended this work to a large order $l$ (but not necessarily infinite) and to any real argument $\beta|J|$ varying from a finite value to infinity. The study of the Bessel differential equation in the large $l$-limit necessitates the introduction of the dimensionless auxiliary variables:

$$\zeta = -\frac{J}{|J|}\left[\sqrt{1+z^2} + \ln\left(\frac{|z|}{1+\sqrt{1+z^2}}\right)\right], \quad |z| = \frac{\beta|J|}{l} \tag{56a}$$

which corresponds to the following transform of the argument of the Bessel function

$$\beta|J| = |z|l \approx |\zeta|l, \text{ as } T \to T_c = 0 \text{ K}. \tag{56b}$$

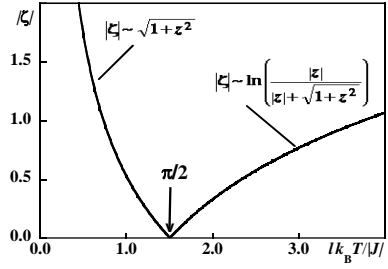

**Figure 6.** Thermal variations of $|\zeta|$ given by equation (56a), for various values of $lk_BT/|J| = 1/|z|$.

Thus $|\zeta|$ plays the role of the inverse of an effective dimensionless coupling near $T_c$. The correspondence with the writing of Chakravarty *et al.* is $|\zeta| = t^{-1}$ [6]. The numerical study of $|\zeta|$ is reported in Fig. 6. We observe that there are two branches. $|\zeta|$ vanishes for a numerical value of $|z_0|^{-1}$ very close to $\pi/2$ so that there are 3 domains which will be physically interpreted in next subsection. Let $T_0$ be the corresponding temperature

$$l\frac{k_BT_0}{|J|} = \frac{\pi}{2}. \tag{57}$$

In the formalism of renormalization group $T_0$ is called a *fixed point*. In the present $2d$ case we have $l \to +\infty$. We then derive that $T_0 \to T_c = 0$ K as $l \to +\infty$ so that the critical temperature can be seen as a *non trivial* fixed point. In other words it means that all the thermodynamic functions can be expanded as series of current term $|T - T_0|$ near $T_0 \approx T_c = 0$ K, in the infinite $l$-limit. Finally Fig. 6 is nothing but the low-temperature diagram of magnetic phases and $|\zeta|$ defined by equation (56a) gives the analytic expression of branches.

For convenience, we introduce the dimensionless coupling constant $g$ at temperature $T$ as well as its reduced value $\bar{g}$:

$$g = \frac{k_BT}{|J|}, \quad \bar{g} = \frac{T}{T_c}. \tag{58}$$

$g$ measures the strength of spin fluctuations. $\bar{g}$ is a universal parameter and is $l$-independent. At the critical point $T_0 = T_c$ we have $\bar{g} = 1$. Owing to equation (57) the critical coupling $g_c$ can be written as:

$$g_c = \frac{k_BT_c}{|J|}; \quad g_c = \frac{\pi}{2l} \text{ or } g_c l = \frac{\pi}{2}. \tag{59}$$

Chubukov *et al.* have found that, at the critical temperature $T_c$, the critical coupling is $g_c = 4\pi/\Lambda$ where $\Lambda = 2\pi/a$ is a relativistic cutoff parameter ($a$ being the lattice spacing) [10]. Thus $\Lambda^{-1}$ appears as a length scale. Haldane has evaluated $g_c$ in the case of a classical spin lattice [11,12]. He proposed $g_c^H = 2\sqrt{d}\,a/S$ or equivalently $g_c^H = 2a/S$ if referring to the vertical rows or horizontal lines of the $2d$-lattice characterized by the same exchange energy $J = J_1 = J_2$. In our case $S = 1$ so that if comparing both results for $g_c$

$$g_c^H = 2a \; ; \; g_c = \frac{4\pi}{\Lambda} \text{ or } g_c\Lambda = 4\pi, \; \Lambda = \frac{2\pi}{a}. \tag{60a}$$

At this step the question is: how to relate the results given by equations (59) and (60a) i.e., $g_c l = \pi/2$ and $g_c \Lambda = 4\pi$?

In the first case $g_c l = \pi/2$ is obtained by a pure numerical method because this is the zero of the function $|\zeta| = f(1/|z|)$ given by equation (56a) with $1/|z| = gl$. The second case $g_c\Lambda = 4\pi$ supposes to take into account the volumic density of $g_c$. According to Chubukov *et al.*, for $D = 3$, $g_c$ is such as $g_c^{-1} = (2\pi)^{-3} \int P^{-2} d^3P$ (for $D = 3$) where $\boldsymbol{P} = (\hbar\boldsymbol{k}, \hbar\omega/c)$ is the relativistic momentum associated with the spin wave of wave vector $\boldsymbol{k}$ and energy $\hbar\omega$.

First we examine the problem of volumic density. As we focus on the static aspect i.e., the volume available to the spin momentum, the relativistic momentum $\boldsymbol{P}/\hbar$ reduces to $\boldsymbol{S}$ (in $\hbar$-unit). As a result the extremity of the classical spin can sweep the surface of a $(D-1)$-dimensional sphere i.e., a $d$-sphere. Thus the elementary volume is $d^dS = s_d S^{d-1} dS$ where $s_d = 2\pi^{d/2}/\Gamma(d/2)$ is the surface of the $d$-sphere. If referring $d^dS$ per unit angle we have $d^dS/(2\pi)^d = [s_d/(2\pi)^d] S^{d-1} dS$. Finally we must take into account the multiplicity of the spin $2S + 1 \sim 2S$ as $S \gg 1$ so that the final elementary volume per degree of multiplicity is $d^dS/2S(2\pi)^d$ i.e., $dV_S = [s_d/2(2\pi)^d] S^{d-2} dS$. Due to our conventional writing $S$ varies between 0 and unity; the integration gives $V_S = s_d/2(2\pi)^d(d-1)$ and

$$g_c = V_S^{-1} = \frac{2(d-1)}{K_d}, \; K_d^{-1} = \left[\frac{s_d}{(2\pi)^d}\right]^{-1} = 2^{d-1}\pi^{d/2}\Gamma(d/2). \tag{60b}$$

This is precisely the result derived by Chakravarty *et al.* as the mathematical solution of the recursion relations established between $g$ and $t = |\zeta|^{-1}$ through a one-loop renormalization process [6]. Thus $g_c = 0$ when $d = 1$ as expected and $g_c = 4\pi$ when $d = 2$.

Second, for obtaining a relation between $l$ and $\Lambda$, we must also consider the volume of the lattice unit cell of spacing $a$. For sake of simplicity we restrict to the case $D = 3$. In equation (59) we introduce the density of exchange energy $|J|S(S+1)/a^2 \sim |J|S^2/a^2 = |J|/a^2$ as $S = 1$ in our conventional writing. In the associated $D$-space-time ($D = 3$) the volume of the phase space is $V_\phi = V_S a^3 = a^3/4\pi$. It means that the value of $g$ per phase space volume is $g/V_\phi = (4\pi/a^3)g$.

As a result $g = g_c l$ given by equation (59) becomes $g = (4\pi/a^3).l k_B T/(|J|/a^2) \to (4\pi l/a).k_B T_c/|J|$ i.e., $g = 2l\Lambda g_c = g_c^* \Lambda^*$ as $T_0 \to T_c = 0$ K so that the new scale is

$$\Lambda^* = 2l\Lambda \gg \Lambda \text{ as } l \to +\infty; \; \Lambda^* = \frac{2\pi}{a^*}, \; 2l = \frac{\Lambda^*}{\Lambda} = \frac{a}{a^*}. \tag{60c}$$

*Thus the l-index of the Bessel functions appears as the ratio of two different scales of reference $\Lambda$ and $\Lambda^*$, respectively associated with the lattices of spacing $a$ and $a^* = a/2l \ll a$.*

We introduce:

(i) the thermal de Broglie wavelength $\lambda_{DB}$;

(ii) the low-temperature spin wave celerity $c = 2\sqrt{2}|J|Sa/\hbar$ along the diagonal of the lattice of spacing $a$ if $J = J_1 = J_2$, with $\sqrt{S(S+1)} = 1$; if $J_1 \neq J_2$ the celerity components become $c_i = 2|J_i|a/\hbar$ along the lattice horizontal lines ($i = 1$, $x$-axis) or the vertical ones ($i = 2$, $y$-axis) and the propagation axis shows the angle $\alpha$ with respect to the $x$-axis such as $\tan\alpha = c_y/c_x = |J_2/J_1|$;

(iii) the slab thickness $L_\tau$ of the $D$-space-time along the $i\tau$-axis ($D = 3$):

$$\lambda_{DB} = 2\pi L_\tau, \; L_\tau = \frac{\hbar c}{k_B T}. \tag{61}$$

By definition we must have

$$\lambda_{DB} \gg a \tag{62}$$

i.e., $\Lambda = 2\pi/a \gg 2\pi/\lambda_{DB}$ or equivalently with $\Lambda^* = 2l\Lambda$

$$\Lambda^* \gg \Lambda \gg L_\tau^{-1} \text{ or } \Lambda^{*-1} \ll \Lambda^{-1} \ll L_\tau. \tag{63}$$

Thus $\Lambda^{-1}$ (or $\Lambda^{*-1}$) appears as a short distance cutoff. No such intrinsic cutoff exists for the imaginary variable $\tau$. At the critical point $T_c = 0$ K $\lambda_{DB} \to +\infty$ (as well as $L_\tau$) and spins are strongly correlated. For $T > T_c$ $\lambda_{DB}$ and $L_\tau$ become finite and diminish as the spin-spin correlation magnitude when $T$ increases. The adequate tool for estimating this correlation between any couple of spins is the correlation length $\xi$. As a result $\lambda_{DB}$ (or $L_\tau$) appears as the good unit length for measuring $\xi$.

Under these conditions we generalize the application of the cutoff parameter $\Lambda$. $|z|$ defined by the second of equation (56a) can be rewritten if using equations (58) and (60c)

$$|z^*| = \frac{\beta|J|}{\Lambda^*} = \frac{z_c^*}{g}, \quad z_c^* = \frac{1}{4\pi}, \quad \Lambda^* = 2l\Lambda \tag{64}$$

and, for a lattice such as $J = J_1 = J_2$, $|z^*|/\Lambda^* = |z|l = \beta|J|$ finally appears as

$$|z^*|/\Lambda^* = \beta|J| = \frac{L_\tau}{2a\sqrt{2}} \tag{65a}$$

owing to the relation $\hbar c = 2\sqrt{2}|J|Sa$ (with $S = 1$). For a general lattice such as $J_1 \neq J_2$ one can use the relation $\hbar c_u = 2|J_i|Sa$ where $c_u$ is the spin-wave celerity along the $x$- ($u = x$, $i = 1$) or the $y$-axis ($u = y$, $i = 2$). As a result the previous equation is slightly modified. Finally the ratio $\Lambda^*/\beta = \Lambda^* k_B T$ can be seen as a new temperature scale.

If examining equation (65a) and noting that $\beta|J| \approx |\zeta| \approx \zeta^*/\Lambda^*$ near $T_c = 0$ K (cf equation (56b) in the $\Lambda$- or in the $\Lambda^*$-scale) with $g_c^* = 2a\sqrt{2}$ (along the diagonal of the lattice) we can write $2a\sqrt{2}\beta|J| \approx g_c^*/\zeta^*/\Lambda^* = L_\tau$ i.e., with $g = g_c^*\Lambda^*$ in the $\Lambda^*$-scale

$$g/\zeta^*/\Lambda^* \approx L_\tau \Lambda^*. \tag{65b}$$

$L_\tau \Lambda^* = 2\lambda_{DB}l/a = \lambda_{DB}/a^*$ (cf equation (60c)) is the dimensionless slab thickness of the $D$-space-time (with $D = d + 1$) in the time-like direction. *We retrieve the result found by Chakravarty et al. i.e., $g/t = g/\zeta = \beta\hbar c$ when establishing the recursion relations between $g$ and $t$ through a one-loop renormalization process, thus allowing to analyze the equilibrium magnetic properties of the 2d-nonlinear $\sigma$ model* [6].

The result given by equation (65b) is universal near $T_c$ so that at the critical point we can write $g_c/t_c = g_c/\zeta_c = \hbar c/k_B T_c$. Owing to equation (60b) and the fact that, when $d = 2$ ($D = 3$) $|\zeta_c|$ diverges ($t_c = 0$) when $g$ tends to $T_c = 0$ K. The unique solution for $t_c$ is

$$t_c = |\zeta_c|^{-1} = \frac{d-2}{K_d}, \quad K_d^{-1} = 2^{d-1}\pi^{d/2}\Gamma(d/2). \tag{65c}$$

Thus we again retrieve the result of *Chakravarty et al.* [6]. It means that, when $d > 2$, $t_c$ and $|\zeta_c|$ become finite.

Finally, if considering the correlation length as a scaling parameter near the critical point $T_c = 0$ K, its measure $\xi$ along the diagonal of the lattice (if $J = J_1 = J_2$) characterized by a spacing $a$ (i.e., in the $\Lambda$-scale) is $\xi_\Lambda = \xi 2a\sqrt{2}$ (or $\xi_\Lambda\Lambda = \xi 4\pi\sqrt{2}$) and $\xi_\tau = \xi L_\tau$ along the slab thickness of the $D$-space-time (i.e., the $i\tau$-axis), due to *scale invariance*. These respective notations can be generalized to a lattice such as $J_1 \neq J_2$ and for any renormalizable physical parameter. As a result we have the dimensionless relations near $T_c = 0$ K

$$\xi = \frac{\xi_\Lambda}{2a\sqrt{2}} = \frac{\xi_\tau}{L_\tau}, \quad \frac{\xi_\tau}{\xi_\Lambda} = \frac{L_\tau}{2a\sqrt{2}} = \beta|J|, \; J = J_1 = J_2, \text{ as } T \to T_c = 0,$$

$$\xi_u = \frac{\xi_{\Lambda,u}}{2a} = \frac{\xi_\tau}{L_\tau}, \quad \frac{\xi_\tau}{\xi_{\Lambda,u}} = \frac{L_\tau}{2a} = \beta|J_i|, \; J_1 \neq J_2 \; (u = x, i = 1; u = y, i = 2), \text{ as } T \to T_c = 0. \tag{66}$$

### 4.2 Low-temperature behaviours of the spin-spin correlation $<S_{0,0} \cdot S_{k,k'}>$

As previously seen (*cf* equations (48a), (48b)) the spin-spin correlation is expressed owing to the ratios $P_{l\pm1} \approx \lambda_{l\pm1}(zl)/\lambda_l(zl) = I_{l\pm1}(zl)/I_l(zl)$ near the critical point $T_c = 0$ K. As the argument $zl$ is replaced by $z^*\Lambda^*$ it is easy to show that, owing to the behaviour of Bessel functions when $l \to +\infty$ ($\Lambda^* \to +\infty$), $\lambda_{l\pm1}(zl)/\lambda_l(zl) \sim \lambda_{\Lambda^*\pm1}(z^*\Lambda^*)/\lambda_{\Lambda^*}(z^*\Lambda^*)$ i.e., $P_{l\pm1} \sim P_{\Lambda^*\pm1}$ as $T \to T_c = 0$ (see Appendix A.3). Simultaneously $|\zeta|/l$ must be replaced by $\zeta^*/\Lambda^*$.

At first sight all these quantities seem to strongly depend on $\Lambda^*$ whereas they must show a universal behaviour near the critical point. As a result scaling parameters i.e., parameters which are $\Lambda^*$-independent must be introduced so that the spin-spin correlation as well as the correlation length are scale-independent near $T_c = 0$ K, as expected.

Chakravarty *et al.* [6] have introduced the physical parameters $\rho_s$ and $\Delta$ defined as:

$$\rho_s = |J|(1-\bar{g}), \quad \Delta = |J|(\bar{g}-1). \tag{67}$$

In the 2*d*-case $\rho_s$ and $\Delta$ have the dimension of an energy $JS^2$ (in our case $J$). $\rho_s$ is the spin stiffness of the ordered ground state (Néel state for an antiferromagnet) and $\Delta$ is the $T=0$-energy gap between the ground state and the first excited state. In the framework of the classical spin approximation the spectrum is quasi continuous. In our case it means that $\Delta$ is very small.

At the critical point $T_c = 0$ K $\bar{g} = 1$: $\rho_s$ and $\Delta$ vanish and, near critically, we have $\rho_s \ll |J|$ and $\Delta \ll |J|$ where $|J|$ finally appears as the bare value of $\rho_s$ and $\Delta$ i.e., their value at 0 K. For all the previous reasons we are then led to introduce the following parameters:

$$\frac{\rho_s}{k_B T} = \frac{1}{g} - \frac{1}{g_c} \quad (T < T_c), \quad \frac{\Delta}{k_B T} = 4\pi \left( \frac{1}{g_c} - \frac{1}{g} \right) \quad (T > T_c) \tag{68}$$

where the factor $4\pi$ appears in $\Delta$ for notational convenience. As a result we can define the following scaling parameters:

$$x_1 = \frac{k_B T}{2\pi \rho_s}, \quad x_2 = \frac{k_B T}{\Delta} \tag{69}$$

where the factor $2\pi$ also appears for notational convenience. As $\rho_s$ and $\Delta$ vanish at $T_0 = T_c$, $x_1$ and $x_2$ become infinite at this fixed point. They are scaling parameters as well as $|z^*|/z_c^*$ and $\zeta^*/\Lambda^*$ (see Appendix A.2). From a physical point of view and as noted by Chakravarty *et al.* [6] as well as by Chubukov *et al.* [10], these parameters control the scaling properties of the magnetic system.

There is an analytical continuity between $x_1$ and $x_2$ when $T_0 = T_c$. As a result there are only 3 domains of predominance: $x_1 \ll 1$ ($T < T_c$ and $|\zeta|/4\pi < 1-\bar{g}$, Zone 1) i.e., $\rho_s \gg k_B T$, $x_2 \ll 1$ ($T > T_c$ and $|\zeta|/4\pi < \bar{g}-1$, Zone 2) i.e., $\Delta \gg k_B T$; finally $x_1 \gg 1$ ($T < T_c$ and $|\zeta|/4\pi > 1-\bar{g}$, Zone 3) i.e., $\rho_s \ll k_B T$ and $x_2 \gg 1$ ($T > T_c$ and $|\zeta|/4\pi > \bar{g}-1$, Zone 4) i.e., $\Delta \ll k_B T$. Along the line $T = T_c$ ($\bar{g} = 1$), we directly reach the *Néel line* (see Fig. 7). Each of these domains previously described corresponds to a particular magnetic regime. The physical meaning of each regime can be derived from the low-temperature study of the ratio $P_{\Lambda^*\pm1} \approx \lambda_{\Lambda^*\pm1}(z^*\Lambda^*)/\lambda_{\Lambda^*}(z^*\Lambda^*)$ *vs* $x_1$ or $x_2$.

The first step consists in expressing $\zeta^*/\Lambda^*$ appearing in $P_{\Lambda^*\pm1}$ as a scaling parameter *vs* $x_1$ or $x_2$ (*cf* equation (A.22)). In Appendix A.2 we have rigorously shown that

$$\zeta^*/\Lambda^* \approx 2\text{arcsinh}\left(\frac{\exp(-1/x_1)}{2}\right), \quad \text{(Zones 1 and 3)}, \tag{70}$$

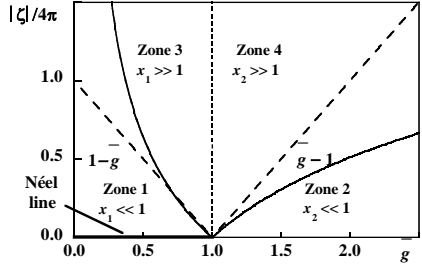 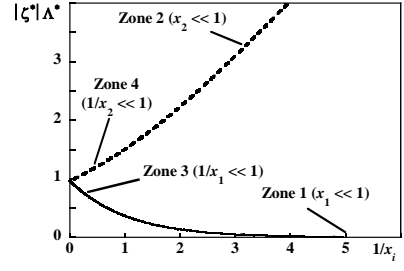

**Figure 7.** Thermal variations of $|\zeta|/4\pi$ vs $\bar{g}$ respectively defined by equations (56a) and (58) and domains of predominance vs dimensionless parameters $x_1$ and $x_2$ defined by equation (69).

**Figure 8.** Plot of $|\zeta^*|/\Lambda^*$ vs the scaling parameters $x_1$ and $x_2$ defined by equation (69).

$$|\zeta^*|/\Lambda^* \approx 2\operatorname{arcsinh}\left(\frac{\exp(1/2x_2)}{2}\right), \text{ (Zones 2 and 4).} \quad (71)$$

The corresponding behaviours are reported in Fig. 8. The asymptotic expansions of $|\zeta^*|/\Lambda^*$ can then be derived for the four zones of the magnetic diagram. We have

$$|\zeta^*|/\Lambda^* \approx \exp(-1/x_1), \; x_1 \ll 1 \text{ (Zone 1)}; \quad (72)$$

$$|\zeta^*|/\Lambda^* \approx \frac{1}{x_2} + 2\exp(-1/x_2), \; x_2 \ll 1 \text{ (Zone 2)}; \quad (73)$$

$$|\zeta^*|/\Lambda^* \approx 2\ln\left(\frac{1+\sqrt{5}}{2}\right) - \frac{2}{\sqrt{5}x_1}, \; x_1 \gg 1 \text{ (Zone 3)}, \quad (74)$$

$$|\zeta^*|/\Lambda^* \approx 2\ln\left(\frac{1+\sqrt{5}}{2}\right) + \frac{1}{\sqrt{5}x_2}, \; x_2 \gg 1 \text{ (Zone 4)}. \quad (75)$$

At the common frontier between Zones 3 and 4, when directly reaching $T_c$, $x_1$ and $x_2$ become infinite and the respective expressions of $|\zeta^*|/\Lambda^*$ show the common limit:

$$C = |\zeta_c^*|/\Lambda^* = 2\ln\left(\frac{1+\sqrt{5}}{2}\right) = 0.962\;424, \; x_1 \to +\infty, x_2 \to +\infty. \quad (76)$$

The ratio $\alpha = (1+\sqrt{5})/2$ is the golden mean. As a result, starting from a closed expression of $|\zeta|$ given by equation (56a) we directly obtained for $|\zeta^*|/\Lambda^*$ the result of Chubukov *et al.* derived from a renormalization technique and called $X_i(x_i)$, $i = 1,2$ [10].

Consequently, in a second step, the ratio $P_{\Lambda^* \pm 1} \approx \lambda_{\Lambda^* \pm 1}(z^*\Lambda^*)/\lambda_{\Lambda^*}(z^*\Lambda^*)$ can be expressed vs $x_1$ or $x_2$ near $T_c = 0$ K as well as the spin-spin correlation $<S_{0,0}.S_{k,k'}> = 3<S_{0,0}^z.S_{k,k'}^z>$ (*cf* equations (32) and (48)). This work is detailed in Appendix *A.3* where it has appeared that, for physical reasons explained at the end of this appendix, $P_{\Lambda^* \pm 1}$ must be renormalized (see Appendix *A.4*).

If $\tilde{P}_{\tilde{\Lambda} \pm 1}$ is the renormalized expression of $P_{\Lambda^* \pm 1}$, with the condition $\tilde{P}_{\tilde{\Lambda} \pm 1} \to 1$ as $T \to T_c = 0$ K, the low-temperature renormalized spin-spin correlation can be written as the following asymptotic limit

$$<\widetilde{S_{0,0}.S_{k,k'}}> \approx <\widetilde{S_{0,0}.S_{k,k'}}>_{\tilde{\Lambda}} \approx \frac{1}{2}\left[\tilde{P}_{\tilde{\Lambda}+1}^{k+k'} + \tilde{P}_{\tilde{\Lambda}-1}^{k+k'} + ...\right], \text{ as } T \to 0, \tilde{\Lambda} \to +\infty \quad (77)$$

with the correspondence $\beta|J|=|z^*|/\Lambda^*=|\tilde{z}|\tilde{\Lambda}$ ($\tilde{\Lambda} = \alpha\Lambda^*$). The low-temperature expressions of $\tilde{P}_{\tilde{\Lambda}\pm 1}$ are given by equation (A.25) for the zones 1 to 4 of the magnetic phase diagram. We have shown in Appendix A.4 that the key renormalized spin-spin correlation between first-nearest neighbours $<\tilde{S}_{0,0}.\tilde{S}_{0,1}>$ can be put under the following form if $J = J_1 = J_2$

$$<\tilde{S}_{0,0}.\tilde{S}_{0,1}>\approx <\tilde{S}_{0,0}.\tilde{S}_{0,1}>_{\tilde{\Lambda}} \approx \frac{1}{2}\left[\tilde{P}_{\tilde{\Lambda}+1}+\tilde{P}_{\tilde{\Lambda}-1}+...\right]\approx -\frac{J}{|J|}\frac{d\ln\lambda_{\tilde{\Lambda}}(|\tilde{z}|/\tilde{\Lambda})}{d(|\tilde{z}|/\tilde{\Lambda})}, \text{ as } T \to 0 \qquad (78a)$$

where $\lambda_{\tilde{\Lambda}}(|\tilde{z}|/\tilde{\Lambda})$ is the dominant eigenvalue in the infinite $\tilde{\Lambda}$-limit or equivalently

$$<\tilde{S}_{0,0}.\tilde{S}_{01}>\approx -\frac{J}{|J|}\left[1-f(x_i)+...\right], \text{ as } T \to 0, \qquad (78b)$$

with

$$f(x_1)=\frac{8\pi}{e}|\tilde{\zeta}_1|\tilde{\Lambda}_1\left(1-\frac{x_1}{2}\right), |\tilde{\zeta}_1|\tilde{\Lambda}_1 \approx |\zeta^*|/\Lambda^*, \text{ Zone 1 } (x_1 << 1), \text{ as } T \to 0,$$

$$f(x_2)=|\tilde{\zeta}_2|\tilde{\Lambda}_2 \approx |\zeta^*|/\Lambda^*, \text{ Zone 2 } (x_2 << 1), \text{ as } T \to 0,$$

$$f(x_3)=|\tilde{\zeta}_3|\tilde{\Lambda}_3 \approx |\zeta^*|/\Lambda^*, \text{ Zone 3 } (x_1 >> 1), \text{ Zone 4 } (x_2 >> 1), \text{ as } T \to 0, \qquad (79)$$

due to the fact that $|\zeta^*|/\Lambda^*$ is a scaling parameter given by equations (72)-(75) and where $\tilde{\Lambda}_1, \tilde{\Lambda}_2$ and $\tilde{\Lambda}_3$ are defined in equation (A.25).

Thus, in Zone 1 ($x_1 << 1$), $f(x_1) \to 0$ as $T \to 0$ and $|<S_{0,0}.S_{0,1}>| \to 1$. In Zone 2 ($x_2 << 1$), due to equation (73) $|\tilde{\zeta}_2|\tilde{\Lambda}_2 \approx x_2|\zeta^*|\Lambda^* \approx 1+2x_2\exp(-1/x_2)\to 1$, $f(x_2) \to 1$ as $T \to 0$ and $|<S_{0,0}.S_{0,1}>| \approx x_2\exp(-1/x_2) \to 0$. We tend towards an assembly of non-correlated spins. In Zones 3 ($x_1 >> 1$) and 4 ($x_2 >> 1$) $|<S_{0,0}.S_{0,1}>| \to 1$ as in Zone 1. As a result this is the low-temperature behaviour of the correlation length $\xi$ which is going to allow the characterization of the magnetic order nature.

### 4.3 Low-temperature behaviours of the correlation length

If using the expression of the correlation length given by equation (51a) as well as the scale invariance property near $T_c = 0$ K given by equation (66) the measure $\xi$ of $\xi_x$ and $\xi_\tau$ is such as

$$\xi=\frac{\xi_{\Lambda^*,x}}{2a^*}\approx \frac{\xi_{\Lambda,x}}{2a}=\frac{\xi_x}{2a}\approx \frac{\xi_\tau}{\hbar c/k_B T}=(f(x_i))^{-1}=(|\zeta^*|/\Lambda^*)^{-1}, i = 1,3, \text{ as } T \to 0 \qquad (80)$$

where $a$ is the lattice spacing as well as $a^*=a/2l$. It means that we can immediately derive the low-temperature correlation length $\xi$.

Recalling that $x_1 = k_B T/2\pi\rho_s$ where $\rho_s$ is the spin stiffness, the lattice spacing $a$ is such as $a = \hbar c/2|J|$ with $\rho_s \approx |J|$ as $\bar{g} = T/T_c$ vanishes with $T$ near $T_c = 0$ K. For a lattice of spacing $a$ we finally derive in Zone 1 owing to equations (79) and (80):

$$\xi_x = \frac{e}{8}\frac{\hbar c}{2\pi\rho_s}\exp\left(\frac{2\pi\rho_s}{k_B T}\right)\left(1+\frac{k_B T}{4\pi\rho_s}\right), x_1 << 1 \text{ (Zone 1)}. \qquad (81)$$

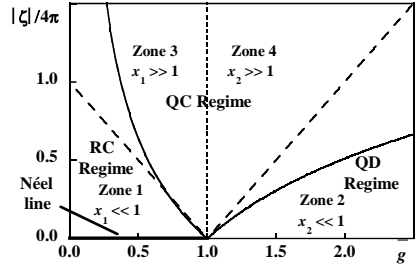

**Figure 9.** Magnetic regime for each domain of predominance of $|\zeta|/4\pi$ vs $\bar{g}$ respectively defined by equations (56a) and (58); the abbreviations stand for Renormalized Classical (RC), Quantum Critical (QC) and Quantum Disordered (QD) regimes.

We exactly retrieve the result first obtained by Hasenfratz and Niedermayer [15] and confirmed by Chubukov *et al.* [10]. This characterizes the *Renormalized Classical Regime* (RCR) for which $\rho_s \gg k_B T$: the divergence of $\xi$ describes a *long-range order* when $T$ approaches $T_c = 0$ K. Spins are aligned ($J < 0$) or antialigned ($J > 0$) inside quasi rigid quasi independent Kadanoff square blocks of side $\xi$ if $J = J_1 = J_2$.

In Zone 2 ($x_2 \ll 1$) where $x_2 = k_B T/\Delta$ we have owing to equations (79) and (80):

$$\xi_\tau \approx \frac{\hbar c}{\Delta}, \, x_2 \ll 1 \text{ (Zone 2)} \tag{82}$$

We deal with the *Quantum Disordered Regime* (QDR) characterized by $\Delta \gg k_B T$. Owing to equation (69) we have $\Delta = k_B T/x_2$ so that $\xi_\tau \approx L_\tau x_2 \ll L_\tau$ as $x_2 \ll 1$. Equivalently, due to the fact that $L_\tau = \hbar c / k_B T$ and $a = \hbar c / 2|J|$ we have $\xi_x \approx 2ax_2 \ll 2a$: we then pass from no $T=0$-order to a *short-range order* when $T$ increases. The magnetic structure is made of spin dimers or aggregates of spin dimers organized in Kadanoff blocks of small size $\xi_\tau$ that we can assimilate to blobs weakly interacting between each others. We deal with a *spin-fluid*. The detailed study is out of the framework of the present article. From a formal point of view it is often phrased in term of Resonating Valence Bonds (RVB) between pairs of quantum spins (considered here in the classical spin approximation) [29]-[31].

In Zones 3 ($x_1 \gg 1$) and 4 ($x_2 \gg 1$) we have

$$\xi_\tau \approx C^{-1} \frac{\hbar c}{k_B T} \left(1 + \frac{2}{C\sqrt{5}x_1}\right), \, x_1 \gg 1 \text{ (Zone 3)},$$

$$\xi_\tau \approx C^{-1} \frac{\hbar c}{k_B T} \left(1 - \frac{1}{C\sqrt{5}x_2}\right), \, x_2 \gg 1 \text{ (Zone 4)} \tag{83}$$

where $C$ is given by equation (76). We now deal with the *Quantum Critical Regime* (QCR). In Zone 3 we have $\rho_s \ll k_B T$ whereas in Zone 4 $\Delta \ll k_B T$. The divergence of $\xi$ describes a *medium-range order* when $T$ approaches $T_c = 0$ K. Spins are aligned ($J < 0$) or antialigned ($J > 0$) inside quasi rigid quasi independent Kadanoff square blocks of side $\xi$ if $J = J_1 = J_2$. But, if comparing with the *Renormalized Classical Regime* (RCR) and the *Quantum Disordered* one (QDR), we have $\xi_{QDR} < \xi_{QCR} < \xi_{RCR}$. Thus Kadanoff blocks show a smaller size when passing from Zone 1 to Zone 2 through Zones 3 and 4.

As a result each behaviour of the correlation length characterizes a magnetic regime. All the predominance domains of these regimes are summarized in Fig. 9.

At the frontier between Zones 3 ($\rho_s \ll k_B T$) and 4 ($\Delta \ll k_B T$) i.e., along the vertical line reaching the Néel line at $T_c$, $x_1$ and $x_2$ become infinite so that:

$$\xi_\tau \approx C^{-1} \frac{\hbar c}{k_B T}, \, T = T_c. \tag{84}$$

i.e., $\xi_\tau \approx L_\tau$ as $C^{-1}$ is close to unity, as predicted by the renormalization group analysis [6,10]. As $\xi_\tau$ diverges according to a $T^{-1}$-law the critical exponent is:

$$\nu = 1 \tag{85}$$

in the $D$-space-time. Owing to previous results the correlation length can also be written as

$$\xi_x \approx \frac{1}{1-|<\widetilde{S_{0,0}.S_{0,1}}>|}, \text{ as } T \to 0, \tag{86}$$

where $|<\widetilde{S_{0,0}.S_{0,1}}>|$ is the renormalized spin-spin correlation between first-nearest neighbours (0,0) and (0,1) as $J = J_1 = J_2$, expressed near $T_c = 0$ K. We retrieve the result predicted in equation (52d). If using the expression of $|<\widetilde{S_{0,0}.S_{0,1}}>|$ given by equation (78a) the correlation length $\xi_x$ can also be expressed as

$$\xi_x = \xi_y \approx \frac{1}{1-\dfrac{d\ln\lambda_{\widetilde{\Lambda}}(|\widetilde{z}|/\widetilde{\Lambda})}{d(|\widetilde{z}|/\widetilde{\Lambda})}}, J = J_1 = J_2, \text{ as } T \to 0, \xi = \xi_x\sqrt{2}. \tag{87}$$

When $J_1 \neq J_2$ a similar expression can be derived for $\xi_x$ and $\xi_y$ with here $\xi_x \neq \xi_y$ and $\xi = \sqrt{\xi_x^2 + \xi_y^2}$; in equation (86) $|<\widetilde{S_{0,0}.S_{0,1}}>|$ is replaced by $|<\widetilde{S_{0,0}.S_{1,0}}>|$ in $\xi_y$ and, in equation (87), $|\widetilde{z}|/\widetilde{\Lambda} = \beta|J|$ becomes $|\widetilde{z_i}|/\widetilde{\Lambda} = \beta|J_i|$, $i = 1,2$.

*This result is exclusively valid for 2d magnetic systems characterized by isotropic couplings because the correlation function reduces to the spin-spin correlation.*

As a result it becomes possible to characterize the nature of magnetic ordering owing to the $T$-decreasing law derived from equation (86)

$$|<\widetilde{S_{0,0}.S_{0,1}}>| \approx f_u(T) = 1 - \xi(T)^{-1}, \text{ as } T \to 0 \tag{88a}$$

where $u$ recalls the nature of the magnetic regime: $u$ = RCR (Zone 1), QCR (Zones 3 and 4) and QDR (Zone 2). As previously seen we have $\xi_{QDR} < \xi_{QCR} < \xi_{RCR}$ so that

$$f_{QDR}(T) < f_{QCR}(T) < f_{RCR}(T), \text{ as } T \to 0. \tag{88b}$$

Thus, in Zone 1 (Renormalized Classical Regime), we have a strong long range order in the critical domain whereas in Zones 3 and 4 (Quantum Critical Regime) the magnitude of magnetic order is less strong. In Zone 2 (Quantum Disordered Regime) we deal with a very short magnitude characteristic of a spin fluid.

## 5. Conclusion

In this paper we have presented the exact general theory of the two-dimensional Heisenberg square lattice composed of classical spins. In the thermodynamic limit a numerical study has allowed to select the higher-degree term of the characteristic $l$-polynomial associated with the zero-field partition function $Z_N(0)$. Under these conditions we have derived an exact closed-form expression of $Z_N(0)$ *valid for any temperature*.

A thermal study of the basic $l$-term of $Z_N(0)$ has allowed to point out *thermal crossovers* between two consecutive eigenvalues $\lambda_l(-\beta J)$ and $\lambda_{l+1}(-\beta J)$, for the first time. When $T$ reaches zero, $l \to +\infty$; all the successive dominant eigenvalues become equivalent so that the critical temperature is $T_c = 0$ K. If using a similar method employed for expressing $Z_N(0)$ we have derived an exact expression for the spin-spin correlations and the cor-relation length ξ *valid for any temperature*. They show new thermal crossovers which are similar to those of $Z_N(0)$.

By assuming the $T=0$-limit of the eigenvalues $\lambda_l(-\beta J)$ we have obtained the low-temperature diagram of magnetic phases characterized by three regimes: the *Renormalized Classical Regime* (RCR), the *Quantum Critical Regime* (QCR) and the *Quantum Disordered Regime* (QDR). This diagram is similar to the one derived through a renormalization approach [6,10].

For each of these regimes, if taking the $T=0$-limit of the general closed-form expression of the correlation length ξ *valid for any temperature*, we obtain the same expression as the corresponding one derived through a renormalization process by several authors but *exclusively valid near the critical point* $T_c = 0$ K [6,10]. In addition we retrieve the good value $\nu = 1$ for the critical exponent. All these results bring a strong validation to the closed-form expressions obtained for $Z_N(0)$, the spin-spin correlations and the correlation length, respectively.

*Finally we have shown that, for the first time, the low-temperature correlation length can be simply expressed by means of the renormalized spin-spin correlation between first-nearest neighbours* $<\tilde{S}_{0,0}.\tilde{S}_{0,1}>$ *but also with the derivative of the logarithm of the dominant eigenvalue* $\lambda_l(\beta|J|)$ *with respect to* $\beta|J|$*, in the limit* $l \to +\infty$*, thus justifying the detailed study of* $Z_N(0)$ *in this article.*

*This result is exclusively valid for 2d magnetic systems characterized by isotropic spin-spin couplings.*

**Appendix**

*A.1 Expression of the zero-field partition function in the thermodynamic limit*

For $T > 0$ K, between two consecutive crossover temperatures $T_{l_i,<}$ and $T_{l_i,>}$, we have shown in the main text that, in the thermodynamic limit ($N \to +\infty$), $Z_N(0)$ can be written as $Z_N(0) = (4\pi)^{8N^2}[u_{\max}(T)]^{4N^2}\{1 + S(N,T)\}$ with $S(N,T) = S_1(N,T) + S_2(N,T)$ (*cf* equations (27) and (28)). $u_{\max}$ is the dominant eigenvalue in $[T_{l_i,<}, T_{l_i,>}]$ according to equation (23).

Due to the numerical property of $u_{l_i,l_i}(T)$ and $u_{l_i,l_j}(T)$ ($l_i \neq l_j$), a classification in the decreasing modulus order can be globally written so that $S(N,T)$ has the form

$$S(N,T) = \sum_{k=0}^{+\infty} X_k(N,T), \ 0 < X_k(N,T) < 1, \quad (A.1)$$

with $X_1(N,T) > X_2(N,T) > ... > X_\infty(N,T)$.

Now we artificially share the infinite series $S(N,T)$ into two parts:

$$S(N,T) = S_{k_i}^B(N,T) + S_{k_i}^E(N,T), T \in [T_{l_i,<}, T_{l_i,>}], \quad (A.2)$$

with $S_{k_i}^B(N,T) = \sum_{k=0}^{k_i} X_k(N,T), S_{k_i}^E(N,T) = \sum_{k=k_i}^{+\infty} X_k(N,T) \cdot S_{k_i}^B(N,T)$ and $S_{k_i}^E(N,T)$ are the beginning and the end of $S(N,T)$, respectively.

We have the natural inequalities $S_{k_i}^B(N,T) < S(N,T)$ and $S_{k_i}^E(N,T) < S(N,T)$. As we deal with an infinite (absolutely convergent) series made of positive vanishing current terms $X_k(N,T) < 1$ it is always possible to find a particular value $k_i = k_1$ of the general index $k$ such as:

$$S_{k_1}^B(N,T) = S_{k_1}^E(N,T) = \frac{\varepsilon}{2}, S(N,T) = \varepsilon, 0 < \varepsilon < 1, T \in [T_{l_i,<}, T_{l_i,>}]. \quad (A.3)$$

If increasing $N >> 1$ of $n > 0$ we automatically have

$$S^K_{k_1}(N+n,T) < S^K_{k_1}(N,T) = \frac{\varepsilon}{2}, \; K = B, E, \; T \in [T_{l_i,<}, T_{l_i,>}] \quad (A.4)$$

because the inequality $0 < X_k(N,T) < 1$ imposes $0 < X_k(N+n,T) < X_k(N,T) < 1$. Finally, if calling $S(N+n,T)$ the sum $S^B_{k_1}(N+n,T) + S^E_{k_1}(N+n,T)$ we have $S(N+n,T) < S(N,T) = \varepsilon$, $T \in [T_{l_i,<}, T_{l_i,>}]$. As a result we derive

$$S(N,T) = S_1(N,T) + S_2(N,T) \to 0, \text{ as } N \to +\infty, \text{ for } T \in [T_{l_i,<}, T_{l_i,>}], \quad (A.5)$$

and due to equation (27) $Z_N(0) \approx (4\pi)^{8N^2}[u_{\max}(T)]^{4N^2}$, as $N \to +\infty$, for any $T \in [T_{l_i,<}, T_{l_i,>}]$.

This reasoning can be repeated for each new range of temperature $[T_{l_j,<}, T_{l_j,>}]$, with $j \neq i$.

In addition, for any predominance range $[T_{l_j,<}, T_{l_j,>}]$, if comparing the current terms $|u_{l_i,l_i}(T)| = F(l,l,0)\lambda_l(\beta|J_1|)\lambda_l(\beta|J_2|)$ and $|u_{l_i,l_j}(T)| = F(l_i,l_j,m)\lambda_{l_i}(\beta|J_1|)\lambda_{l_i}(\beta|J_2|)$ of $S_1(N,T)$ and $S_1(N,T)$ given by equation (28), it is always have possible to find $|u_{l_i,l_i}(T)| > |u_{l_i,l_j}(T)|$ for $l_i = l \leq l_j$ (cf Fig. 2b).

Consequently, if summing these terms over all the ranges $[T_{l_i,<}, T_{l_i,>}]$ so that $T = \sum_{i=0}^{i_{\max}}(T_{l_i,>} - T_{l_i,<})$, $T_{l_0,<} = 0$, $T_{l_{i-1},>} = T_{l_i,<}$ $(i \neq 0)$ and $T_{l_{i_{\max}},>} = T$ we always have

$$S_1(N,T) = \sum_{l=0}^{+\infty}\left[\frac{u_{l,l}(T)}{u_{\max}}\right]^{4N^2} \gg S_2(N,T) = \prod_{i=-(N-1)}^{N}\prod_{j=-(N-1)}^{N}\sum_{l_i=0}^{+\infty}\sum_{l_j=0, l_j \neq l_i}^{+\infty}\frac{u_{l_i,l_j}(T)}{u_{\max}},$$

$$\text{as } N \to +\infty \quad (A.6)$$

due to the fact that $1 > |u_{l_i,l_i}(T)/u_{\max}| \gg |u_{l_i,l_j}(T)/u_{\max}| > 0$. As a result

$$Z_N(0) \approx (4\pi)^{8N^2}\sum_{l=0}^{+\infty}{}^t\left[F_{l,l}\lambda_l(-\beta J_1)\lambda_l(-\beta J_2)\right]^{4N^2}, \text{ as } N \to +\infty. \quad (A.7)$$

### A.2 Calculation of $|\zeta^*|\Lambda^*$ near the critical point

In the thermodynamic limit each expression of the thermodynamic functions (spin-spin correlations, correlation length...) involves ratios of Bessel functions $I_{\Lambda^*}(z^*\Lambda^*)$. These functions have to be evaluated in the double limit $\beta|J| = |z^*|\Lambda^* \to +\infty$, $\Lambda^* \to +\infty$. In that case Olver has shown [27] that the argument $\beta|J| = |z^*|\Lambda^*$ must be replaced by $|\zeta^*|\Lambda^*$ where

$$\zeta^* = -\frac{J}{|J|}\left[\sqrt{1+z^{*2}} + \ln\left(\frac{|z^*|}{1+\sqrt{1+z^{*2}}}\right)\right], \; z^* = \frac{\beta J}{\Lambda^*}. \quad (A.8)$$

At the fixed point $z^*_c = 1/4\pi$ we exactly have $|\zeta^*| = 0$. Near this critical point (see Fig. 6), $\sqrt{1+z^{*2}} \approx |\ln(|z^*|/[1+\sqrt{1+z^{*2}}])|$ for any Zone 1 to 4. As a result equation (A.8) reduces to $|\zeta^*| \approx |\ln(|z^*|^{-1} + \sqrt{1+z^{*-2}})|$ or equivalently $|\zeta^*| \approx |\text{arcsinh}(|z^*|^{-1})|$ as $|z^*| \to z^*_c$.

Near $z^*_c = 1/4\pi$, for $T < T_c$ or $T > T_c$, equation (A.8) can also be written $|\ln(|z^*|/2)|$ which depends on $\Lambda^*$. As the ratio $|z^*|/z^*_c = T_c/T$ is independent of $\Lambda^* \gg 1$ a scaling form of $|\zeta^*|\Lambda^*$ can be $|\zeta^*|\Lambda^* \approx 2\ln(|z^*|/2z^*_c\Lambda^{*/2})$ or $2\ln(|z^*_c|/2z^{*\Lambda^{*/2}})$, as $|z^*| \to z^*_c$. Due to the previous remarks we must have

$$|\zeta^*|/\Lambda^* \approx 2\mathrm{arcsinh}(|z^*|/2z_c^*|^{\Lambda^*/2}) \text{ or } |\zeta^*|/\Lambda^* \approx 2\mathrm{arcsinh}(|z_c^*/2z^*|^{\Lambda^*/2}), \text{ as } |z^*| \to z_c^*. \tag{A.9}$$

If $|\zeta^*|/\Lambda^*$ is a scaling parameter we must show that $(|z^*|/z_c^*)^{\Lambda^*}$ or $(z_c^*/|z^*|)^{\Lambda^*}$ is $\Lambda^*$-independent. In Zones 1 ($x_1 \ll 1$) and 3 ($x_1 \gg 1$), $|z^*| > z_c^*$ so that, from the definition of $\rho_s$ (cf equation (68)), we have $\Lambda^*(|z^*|-z_c^*) = \rho_s/k_BT$. In Zones 2 ($x_2 \ll 1$) and 4 ($x_2 \gg 1$) $|z^*| < z_c^*$. We similarly have from the definition of $\Delta$ $\Lambda^*(z_c^*-|z^*|) = \Delta/4\pi k_BT$. If introducing $x_1$ and $x_2$ given by equation (69):

$$\Lambda^*(|z^*|-z_c^*) = \frac{2z_c^*}{x_1}, \quad \Lambda^*(z_c^*-|z^*|) = \frac{z_c^*}{x_2}, \quad z_c^* = \frac{1}{4\pi}. \tag{A.10}$$

Using the well-known relation $(1 \pm u/\Lambda^*)^{\Lambda^*} = \exp(\pm u)$, as $\Lambda^* \to +\infty$, we derive from equation (A.10) that, near $z_c^*$

$$\left(\frac{|z^*|}{z_c^*}\right)^{\Lambda^*/2} = \exp\left(\frac{1}{x_1}\right), \frac{|z^*|}{z_c^*} = \exp\left(\frac{2}{x_1\Lambda^*}\right); \left(\frac{|z^*|}{z_c^*}\right)^{\Lambda^*/2} = \exp\left(-\frac{1}{2x_2}\right), \frac{|z^*|}{z_c^*} = \exp\left(-\frac{1}{x_2\Lambda^*}\right). \tag{A.11}$$

As $x_1$ and $x_2$ are scaling parameters the ratios $(|z^*|/z_c^*)^{\Lambda^*}$ and $(z_c^*/|z^*|)^{\Lambda^*}$ are themselves scaling parameters as well as $|\zeta^*|/\Lambda^*$ given by equation (A.9).

Due to the behaviour of $|\zeta^*|/\Lambda^*$ (cf Fig. 8), if $|z^*| > z_c^*$ ($T < T_c$) $|\zeta^*|/\Lambda^*$ decreases with $|z^*|$ as $|z^*| \to z_c^*$ like the ratio $(z_c^*/|z^*|)^{\Lambda^*/2} = \exp(-1/x_1)$; if $|z^*| < z_c^*$ ($T > T_c$) $|\zeta^*|/\Lambda^*$ decreases when $|z^*|$ increases like the ratio $(z_c^*/|z^*|)^{\Lambda^*/2} = \exp(1/2x_2)$.

As a result, if taking into account these remarks for the previous equations, we can write

$$|\zeta^*|/\Lambda^* \approx 2\mathrm{arcsinh}\left(\frac{\exp(-1/x_1)}{2}\right) \text{ (Zones 1, 3)}, \quad |\zeta^*|/\Lambda^* \approx 2\mathrm{arcsinh}\left(\frac{\exp(1/2x_2)}{2}\right) \text{ (Zones 2, 4)}. \tag{A.12}$$

The various asymptotic expansions of $|\zeta^*|/\Lambda^*$ are given in the main text.

### A.3 Asymptotic expansions of modified Bessel functions of the first kind of large order; application to the low-temperature spin-spin correlations

The expression of spin-spin correlations involves ratios such as $\lambda_{l\pm 1}(zl)/\lambda_l(zl)$ i.e., $I_{l\pm 1}(zl)/I_l(zl)$ where $I_l(zl)$ is the Bessel function of the first kind, with $z = -\beta J/l$.

In the main text, we have seen that, near $T_c = 0$ K, $l$ must be replaced by $\Lambda^* = 2l\Lambda$ and more generally by any new scale $\Lambda' = \alpha l$, as $l \to +\infty$, with the imposed condition $|z|l = |z^*|/\Lambda^* = |z'|/\Lambda' = \beta|J|$. We then have $\lambda_l(|z|l) = \lambda_l(|z^*|/\Lambda^*) = \lambda_l(|z'|/\Lambda')$. When $l \to +\infty$ $\lambda_l(|z|l) \approx \lambda_{\Lambda^*}(|z^*|/\Lambda^*) \approx \lambda_{\Lambda'}(|z'|/\Lambda')$. As a result we can write in the new $\Lambda'$-scale

$$\frac{\lambda_{l+1}(|z|l)}{\lambda_l(|z|l)} \approx \frac{\lambda_{\Lambda^*+1}(|z^*|/\Lambda^*)}{\lambda_{\Lambda^*}(|z^*|/\Lambda^*)} \approx \frac{\lambda_{\Lambda'+1}(|z'|/\Lambda')}{\lambda_{\Lambda'}(|z'|/\Lambda')}, \Lambda' = \alpha l \to +\infty, |z'|/\Lambda' \to +\infty \tag{A.13}$$

with a similar relation if $l + 1$ is replaced by $l - 1$.

Then, if using equations (8) and (48b), the recurrence relations between $I_{\Lambda^*}(|z^*|/\Lambda^*)$, $I_{\Lambda^*+1}(|z^*|/\Lambda^*)$ and $I_{\Lambda^*-1}(|z^*|/\Lambda^*)$ and the condition $|z^*|^{-1} \gg (2|z^*|/\Lambda^*)^{-1}$ as $\Lambda^* \to +\infty$, we have

$$P_{\Lambda^*\pm 1} \approx \frac{\lambda_{\Lambda^*\pm 1}(z^*\Lambda^*)}{\lambda_{\Lambda^*}(z^*\Lambda^*)} \approx -\frac{J}{|J|}\left\{\mp\frac{1}{|z^*|} + \frac{I'_{\Lambda^*}(|z^*|/\Lambda^*)}{I_{\Lambda^*}(|z^*|/\Lambda^*)}\right\}, I'_{\Lambda^*}(|z^*|/\Lambda^*) = \frac{dI_{\Lambda^*}(|z^*|/\Lambda^*)}{d(|z^*|/\Lambda^*)}. \quad (A.14)$$

Due to the polynomial structure of the spin-spin correlation $<S_{0,0}\cdot S_{k,k'}> = 3<S^z_{0,0}\cdot S^z_{k,k'}>$ detailed in the main text (*cf* equation (48)), we have the asymptotic behaviour as $T \to T_c = 0$ K (*cf* equation (78a))

$$<S_{0,0}\cdot S_{k,k'}> \approx \frac{3}{2}\left[\left(\frac{I_{\Lambda^*+1}(z^*\Lambda^*)}{I_{\Lambda^*}(z^*\Lambda^*)}\right)^{k+k'} + \left(\frac{I_{\Lambda^*-1}(z^*\Lambda^*)}{I_{\Lambda^*}(z^*\Lambda^*)}\right)^{k+k'} + \ldots\right], \text{ as } |z^*|/\Lambda^* \to +\infty. \quad (A.15)$$

In Zone 1 exclusively, we have $|z^*| \approx z_c^* \exp(2/x_1\Lambda^*)$ (*cf* equation (A.11)) so that $|z^*|^{-1} << z_c^{*-1}$ except if $|z^*|$ is close to $z_c^*$ but, for Zones 2, 3 and 4, $|z^*|^{-1} \approx z_c^{*-1}$. As a result

$$<S_{0,0}\cdot S_{k,k'}> \approx 3\left(-\frac{J}{|J|}\right)^K\left[\left|\frac{I'_{\Lambda^*}(z^*\Lambda^*)}{I_{\Lambda^*}(z^*\Lambda^*)}\right|^K + \ldots\right], K = k + k', \text{ Zone 1 } (x_1 << 1), |z^*| >> z_c^*;$$

$$<S_{0,0}\cdot S_{k,k'}> \approx 3\left(-\frac{J}{|J|}\right)^K \sum_{\nu=0}^{\lfloor K/2 \rfloor}\binom{K}{2\nu}\frac{1}{|z^*|^{K-2\nu}}\left[\left|\frac{I'_{\Lambda^*}(z^*\Lambda^*)}{I_{\Lambda^*}(z^*\Lambda^*)}\right|^{2\nu} + \ldots\right], \text{ as } T \to 0, |z^*| \sim z_c^*,$$

Zone 1 ($x_1 << 1$), Zone 2 ($x_2 << 1$), Zone 3 ($x_1 >> 1$), Zone 4 ($x_2 >> 1$), (A.16)

where $\lfloor K/2 \rfloor$ is the floor function which gives the integer part of $K/2$.

Olver has shown [27] that the Bessel function $I_{\Lambda^*}(|z^*|/\Lambda^*)$ as well as $I'_{\Lambda^*}(|z^*|/\Lambda^*)$ can be expanded as the following series in the double infinite-limit $\Lambda^* \to +\infty$ and $|z^*|/\Lambda^* \to +\infty$

$$I_{\Lambda^*}(|z^*|/\Lambda^*) \approx \frac{(1+z^{*2})^{-1/4}}{\sqrt{2\pi\Lambda^*}}\left\{\exp(|\zeta^*|/\Lambda^*)\sum_{s=0}^{+\infty}\frac{U_s(u^*)}{\Lambda^{*s}} + \exp(-|\zeta^*|/\Lambda^*)\sum_{s=0}^{+\infty}\frac{U_s(u^*)}{(-\Lambda^*)^s}\right\},$$

$$I'_{\Lambda^*}(|z^*|/\Lambda^*) \approx \frac{(1+z^{*2})^{1/4}}{\sqrt{2\pi\Lambda^*}|z^*|}\left\{\exp(|\zeta^*|/\Lambda^*)\sum_{s=0}^{+\infty}\frac{V_s(u^*)}{\Lambda^{*s}} - \exp(-|\zeta^*|/\Lambda^*)\sum_{k=0}^{+\infty}\frac{V_s(u^*)}{(-\Lambda^*)^s}\right\} \quad (A.17)$$

where $\zeta^*/\Lambda^*$ is given by equation (A.12) and $u^* = 1/\sqrt{1+z^{*2}}$ with $|z^*| = \beta|J|/\Lambda^*$. The coefficients $U_s(u^*)$ and $V_s(u^*)$ which are $u^*$-polynomials are detailed in [27,28].

Introducing the previous series in the ratios appearing in equation (A.16) we have

$$\frac{I'_{\Lambda^*}(|z^*|/\Lambda^*)}{I_{\Lambda^*}(|z^*|/\Lambda^*)} \approx \frac{1}{u^*|z^*|}\frac{1+V_+(u^*)}{1+U_+(u^*)}\left[1-\exp(-2|\zeta^*|/\Lambda^*)\left(\frac{1+V_-(u^*)}{1+V_+(u^*)} + \frac{1+U_-(u^*)}{1+U_+(u^*)}\right) + \ldots\right], \quad (A.18a)$$

as $T \to 0$ where each $u^*$-series of equation (A.17) has been written $X_\pm = 1 + X_\pm(u^*)$ with $X_\pm = U_\pm$ or $V_\pm$ and $X_\pm(u^*) = \sum_{s=1}^{+\infty}X_s(u^*)/(\pm\Lambda^*)^s$.

Near $T_c = 0$ K exclusively, as $\zeta^*/\Lambda^*$ is a scaling parameter, we can then define a new scale $\Lambda^{*'}$ such as $2\zeta^*/\Lambda^* = \zeta^{*'}/\Lambda^{*'}$ i.e., $\Lambda^{*'} = 2\Lambda^*$ as $\Lambda^{*'} \to +\infty$. $\zeta^{*'}/\Lambda^{*'} \approx \zeta^*/\Lambda^*$ has been calculated near $z_c^{*'} = 1/4\pi$ in Appendix *A.2*; the corresponding asymptotic expansions are given in the main text (*cf* equations (72)-(75)).

The $u^*$-coefficients of series $U_\pm(u^*)$ and $V_\pm(u^*)$ become $u^{*'}$ and their corresponding series $U_\pm(u^{*'})$ and $V_\pm(u^{*'})$. Similarly $z^*$ becomes $z^{*'}$ but $z_c^{*'} = z_c^* = 1/4\pi$.

Now it is necessary to know the respective values of these series vs $z_c^{*'} = 1/4\pi$. Using the multiplication theorem [28] and the fact that $z_c^{*'}$ is small, we have $I_{\Lambda^{*'}}\left(z_c^{*'}\Lambda^{*'}\right) \sim \Lambda^{*'\Lambda^{*'}} I_{\Lambda^{*'}}\left(z_c^{*'}\right) \sim \Lambda^{*'\Lambda^{*'}} (z_c^{*'}/2)^2/\Lambda^{*'}!$. Then, owing to the well-known Stirling formula giving $\Lambda^{*'}!$ when $\Lambda^{*'}! \gg 1$, it is easy to derive that

$$I_{\Lambda^{*'}}\left(z_c^{*'}\Lambda^{*'}\right) \approx \frac{\exp(\Lambda^{*'})\left(z_c^{*'}/2\right)^{\Lambda^{*'}}}{\sqrt{2\pi\Lambda^{*'}}}. \tag{A.18b}$$

Equation (A.18a) imposes to express the dimensionless quantity $|\zeta^{*'}|$ defined by equation (56a) near $z_c^{*'}$. As $z_c^{*'}$ is small we have $|\zeta^{*'}| \approx 1 + \ln(z_c^{*'}/2)$. In addition, when $\Lambda^{*'} \gg 1$, the second part in the first of equation (A.17) vanishes so that, by identifying the remaining part and equation (A.18b), we have $U_+(z_c^{*'}) \approx (1+z_c^{*'2})^{1/4}$ i.e., $1+(z_c^{*'}/2)^2$.

Finally, owing to the following relation first found by Olver [27] between coefficient $U_s(u^{*'})$ and $V_s(u^{*'})$

$$V_s(u^{*'}) = U_s(u^{*'}) - u^{*'}(1-u^{*'2})\left(\frac{1}{2}U_{s-1}(u^{*'}) + u^{*'}\frac{dU_{s-1}(u^{*'})}{du^{*'}}\right), \quad s \geq 1. \tag{A.18c}$$

it becomes possible to obtain a relation between series $U_+(u^{*'})$ and $V_+(u^{*'})$. Thus, near the fixed point $z_c^{*'}$ and when $\Lambda^{*'} \gg 1$ it is easy to show that

$$\begin{pmatrix} U_+(z_c^{*'}) \\ V_+(z_c^{*'}) \end{pmatrix} \approx 1 \pm \left(\frac{z_c^{*'}}{2}\right)^2; \tag{A.19a}$$

the sign + of the second member (respectively, the sign −) refers to $U_s(u^{*'})$ (respectively, $V_+(u^{*'})$).

A similar reasoning allows to derive the series $U_-(u^{*'})$ and $V_-(u^{*'})$ but the function $I_{\Lambda^{*'}}\left(z_c^{*'}\Lambda^{*'}\right)$ must be replaced by the other Bessel function $K_{\Lambda^{*'}}\left(z_c^{*'}\Lambda^{*'}\right)$. Near the fixed point $z_c^{*'}$ and when $\Lambda^{*'} \gg 1$, we find:

$$\begin{pmatrix} U_-(z_c^{*'}) \\ V_-(z_c^{*'}) \end{pmatrix} \approx \frac{1}{e}\begin{pmatrix} U_+(z_c^{*'}) \\ V_+(z_c^{*'}) \end{pmatrix} \tag{A.19b}$$

with the same conventional writing defined after equation (A.19a).

A detailed study shows that equation (A.18) must be used exclusively for Zone 1. For Zones 2, 3 and 4 it is just necessary to expand the series of equation (A.18). We have

$$\frac{I'_{\Lambda^{*'}}(|z^{*'}|/\Lambda^{*'})}{I_{\Lambda^{*'}}(|z^{*'}|/\Lambda^{*'})} \approx \frac{1}{u^{*'}|z^{*'}|} - \frac{1}{|z^{*'}|/\Lambda^{*'}} - 2\exp\left(-|\zeta^{*'}|/\Lambda^{*'}\right)\left(\frac{1}{u^{*'}|z^{*'}|} - \frac{1}{2|z^{*'}|/\Lambda^{*'}} + ..\right), \text{ as } T \to 0. \tag{A.20}$$

For Zone 1 ($x_1 \ll 1$), we derive from equation (A.10) that $u^{*'}/\Lambda^{*'} \sim 1/|z^{*'}|/\Lambda^{*'} \sim x_1/2 \, z_c^{*'}$. For Zone 2 ($x_2 \ll 1$) $|z^{*'}| \ll z_c^{*'} = 1/4\pi$ and $u^{*'}/\Lambda^{*'} \sim x_2$. In Zones 3 ($x_1 \gg 1$) and 4 ($x_2 \gg 1$), the current term $u^{*'}/\Lambda^{*'} \sim 1/\Lambda^{*'}$ vanishes as $\Lambda^{*'} \to +\infty$.

We skip the intermediate steps which show no difficulties and give the final results:

$$\left|\frac{I'_{\Lambda^{*'}}(z^{*'}\Lambda^{*'})}{I_{\Lambda^{*'}}(z^{*'}\Lambda^{*'})}\right| = 1 - \frac{2}{ez_c^{*'}}\exp(-1/x_1)\left(1-\frac{x_1}{2}\right)+..., \text{ as } T \to 0, \text{ Zone 1 } (x_1 \ll 1),$$

$$\left|\frac{I'_{\Lambda^{*'}}(z^{*'}\Lambda^{*'})}{I_{\Lambda^{*'}}(z^{*'}\Lambda^{*'})}\right| = \frac{1}{x_2 z_c^{*'}}\left[1+x_2-x_2/\zeta^{*'}/\Lambda^{*'}+...\right], \text{ Zone 2 } (x_2 \ll 1), \qquad (A.21)$$

$$\left|\frac{I'_{\Lambda^{*'}}(z^{*'}\Lambda^{*'})}{I_{\Lambda^{*'}}(z^{*'}\Lambda^{*'})}\right| = \frac{1}{z_c^{*'}}\left(2(1+C)e^{-C}-1\right)\left[1-\frac{2e^{-C}}{2(1+C)e^{-C}-1}/\zeta^{*'}/\Lambda^{*'}+...\right], \text{ Zones 3, 4 } (x_1, x_2 \gg 1),$$

Under these conditions, if taking into account all the previous results and remarks, we can write the low-temperature spin-spin correlation

$$<S_{0,0}\cdot S_{k,k'}> \approx \frac{3}{2}\left[P^{k+k'}_{\Lambda^{*'}+1}+P^{k+k'}_{\Lambda^{*'}-1}+...\right], \quad P_{\Lambda^{*'}\pm 1} \approx \frac{\lambda_{\Lambda^{*'}\pm 1}(z^{*'}\Lambda^{*'})}{\lambda_{\Lambda^{*'}}(z^{*'}\Lambda^{*'})}, \text{ as } T \to 0,$$

$$P_{\Lambda^{*'}\pm 1} \approx -\frac{J}{|J|}\left[1-\frac{8\pi}{e}/\zeta^{*'}/\Lambda^{*'}\left(1-\frac{x_1}{2}\right)+...\right], \text{ as } T \to 0, \text{ Zone 1 } (x_1 \ll 1),$$

$$P_{\Lambda^{*'}\pm 1} \approx -\frac{J}{|J|}\frac{1}{x_2 z_c^{*'}}\left[\mp x_2 + 1 + x_2 - x_2/\zeta^{*'}/\Lambda^{*'}+...\right], \text{ Zone 2 } (x_2 \ll 1),$$

$$P_{\Lambda^{*'}\pm 1} \approx -\frac{J}{|J|}\frac{1}{z_c^{*'}}\left(2(1+C)e^{-C}-1\right)\left[\pm\frac{1}{2(1+C)e^{-C}-1}+1-\frac{2e^{-C}}{2(1+C)e^{-C}-1}/\zeta^{*'}/\Lambda^{*'}+...\right],$$

$$\text{Zone 3 } (x_1 \gg 1), \text{ Zone 4 } (x_2 \gg 1), \quad (A.22)$$

where the scaling parameter $/\zeta^{*'}/\Lambda^{*'} \approx \zeta^*/\Lambda^*$ is respectively given by equations (72)-(75). In Zones 1 and 3 $/\zeta^{*'}/\Lambda^{*'} < 1$ near $T_c = 0$ K. In Zones 2 and 4 $/\zeta^{*'}/\Lambda^{*'} > 1$ (see Fig. 8).

We note that except for Zone 1 $P_{\Lambda^{*'}\pm 1}$ does not tend to unity due to the technique used for establishing the low-temperature expansions of the various Bessel functions [27]. As pointed out by Olver [27] these expressions are defined within a numerical factor. As they are expressed with scaling parameters it becomes possible to renormalize them near $T_c$.

## A.4 Renormalized expressions of the low-temperature spin-spin correlations

We finally focus on the renormalization of the low-temperature spin-spin correlations near $T_c = 0$ K. We define a new scale $\tilde{\Lambda}$ such as $\beta|J|=/z^{*'}/\Lambda^{*'}=|\tilde{z}|\tilde{\Lambda}$ with $\tilde{\Lambda}=\alpha\Lambda^{*'}$.

Owing to the multiplication theorem of the functions $I_{\Lambda^{*'}}(\alpha x)$, for finite $x > 0$, $\alpha > 0$, we have $I_{\Lambda^{*'}}(\alpha x) \approx \alpha^{\Lambda^{*'}}I_{\Lambda^{*'}}(x)$ at first order due to the $\Lambda^{*'}$-infinite limit [28]. As a result $P_{\Lambda^{*'}+1} = I_{\Lambda^{*'}+1}(\alpha z^{*'}\Lambda^{*'})/I_{\Lambda^{*'}}(\alpha z^{*'}\Lambda^{*'}) = I_{\Lambda^{*'}+1}(\tilde{z}\tilde{\Lambda})/I_{\Lambda^{*'}}(\tilde{z}\tilde{\Lambda}) \approx \alpha I_{\Lambda^{*'}+1}(z^{*'}\Lambda^{*'})/I_{\Lambda^{*'}}(z^{*'}\Lambda^{*'})$. In the infinite $\Lambda^{*'}$- and $\tilde{\Lambda}$-limits and due to equation (A.13) we finally have $\tilde{P}_{\tilde{\Lambda}+1} \approx \alpha P_{\Lambda^{*'}+1}$.

$P_{\Lambda^{*'}-1}$ can be expressed as $P_{\Lambda^{*'}+1}$ i.e., with the ratio $\left|I'_{\Lambda^{*'}}(z^{*'}\Lambda^{*'})/I_{\Lambda^{*'}}(z^{*'}\Lambda^{*'})\right|$ characterized by the factor $K(u_c^{*'}z_c^{*'})^{-1}$ (cf equations (A.21) and (A.22)).

As a result the dilation factor $\alpha$ for $P_{\Lambda^{*'}\pm 1}$ is such as

$$\frac{K\alpha}{u^{*'}/z^{*'}} = 1, \quad \alpha = \frac{z_c^{*'}}{K\sqrt{1+z_c^{*'2}}} \approx \frac{z_c^{*'}}{K}, \quad \tilde{\Lambda} = \alpha\Lambda^{*'}. \tag{A.23}$$

In the main text we have seen that the spin-spin correlation $<S_{0,0}.S_{0,1}>$ plays a major role. We must have $|<S_{0,0}.S_{0,1}>| = 1$ at $T_c = 0$ K. Thus the renormalization of the first of equation (A.22) finally imposes to have $\tilde{P}_{\tilde{\Lambda}\pm 1} \to 1$. Owing to equations (8) and (A.14) we can define the renormalized spin-spin correlation $<\widetilde{S_{0,0}.S_{0,1}}>$ as

$$<\widetilde{S_{0,0}.S_{0,1}}> \approx \frac{1}{2}\left[\tilde{P}_{\tilde{\Lambda}+1} + \tilde{P}_{\tilde{\Lambda}-1} + ...\right] = -\frac{J}{|J|}\frac{I'_{\tilde{\Lambda}}(|\tilde{z}|/\tilde{\Lambda})}{I_{\tilde{\Lambda}}(|\tilde{z}|/\tilde{\Lambda})} = -\frac{J}{|J|}\frac{d\ln\lambda_{\tilde{\Lambda}}(|\tilde{z}|/\tilde{\Lambda})}{d(|\tilde{z}|/\tilde{\Lambda})}, \text{ as } T \to 0 \tag{A.24}$$

where $\lambda_{\tilde{\Lambda}}(|\tilde{z}|/\tilde{\Lambda})$ is the dominant eigenvalue (in the limit $\tilde{\Lambda} \to +\infty$). In the limit $T \to 0$, as $x_1$ and $x_2$ are scaling parameters, we have

$$\tilde{P}_{\tilde{\Lambda}\pm 1} \approx -\frac{J}{|J|}\left[1 - \frac{8\pi}{e}|\xi_1|\tilde{\Lambda}_1\left(1-\frac{x_1}{2}\right)+...\right], \alpha_1 = 1/3, \tilde{\Lambda}_1 = \Lambda^{*'}, \text{ Zone 1 } (x_1 << 1),$$

$$\tilde{P}_{\tilde{\Lambda}\pm 1} \approx -\frac{J}{|J|}\left[\mp x_2 + 1 + x_2 - |\xi_2|\tilde{\Lambda}_2 + ...\right], \alpha_2 = x_2 z_c^{*'}/3, \tilde{\Lambda}_2 = x_2\Lambda^{*'}, \text{ Zone 2 } (x_2 << 1),$$

$$\tilde{P}_{\tilde{\Lambda}\pm 1} \approx -\frac{J}{|J|}\left[\pm\frac{1}{2(1+C)e^{-C}-1} + 1 - |\xi_3|\tilde{\Lambda}_3 + ...\right], \alpha_3 = \frac{z_c^{*'}}{3(2(1+C)e^{-C}-1)}, \tilde{\Lambda}_3 = \frac{6e^{-C}\alpha_3}{z_c^{*'}}\Lambda^{*'},$$

$$\text{Zone 3 } (x_1 >> 1), \text{ Zone 4 } (x_2 >> 1) \tag{A.25}$$

where $C$ is given by equation (76).

Near $T_c = 0$ K the key renormalized spin-spin correlation $<\widetilde{S_{0,0}.S_{0,1}}>$ given by equation (A.24) can also be written

$$<\widetilde{S_{0,0}.S_{01}}> \approx -\frac{J}{|J|}\left[1 - f(x_i) + ...\right], \text{ as } T \to 0, \tag{A.26}$$

$f(x_1) = \frac{8\pi}{e}|\xi_1|\tilde{\Lambda}_1\left(1-\frac{x_1}{2}\right)$, Zone 1 $(x_1 << 1)$, $f(x_2) = |\xi_2|\tilde{\Lambda}_2$, Zone 2 $(x_2 << 1)$ ;

$f(x_3) = |\xi_3|\tilde{\Lambda}_3$, Zone 3 $(x_1 >> 1)$, Zone 4 $(x_2 >> 1)$ \hfill (A.27)

with $|\xi_i|\tilde{\Lambda}_i \approx |\zeta^{*'}|/\Lambda^{*'} \approx |\zeta^*|/\Lambda^*$ for Zone $i$ ($i = 1$ to 3) as $|\zeta^*|/\Lambda^*$ is a scaling parameter near $T_c$.

### ORCID iDs

Jacques Curély  https://orcid.org/0000-0002-2635-7927